\documentclass[prx,aps,showkeys,showpacs,superscriptaddress,twocolumn,amsmath,amssymb]
{revtex4-1} 

\usepackage{graphicx}
\usepackage{subfigure}
\usepackage{epstopdf, epsfig}
\usepackage{color}							
\usepackage{setspace}
\usepackage{caption}
\usepackage{float}
\usepackage{natbib}
\usepackage{color}

\begin{document}
\title{Interaction network analysis in shear thickening suspensions }
\author{Marcio Gameiro}
\email{gameiro@icmc.usp.br}
\affiliation{Instituto de Ci\^{e}ncias Matem\'{a}ticas e de Computa\c{c}\~{a}o, Universidade de S\~{a}o Paulo,Caixa Postal 668, 13560-970, S\~{a}o Carlos, SP, Brazil}
\author{Abhinendra Singh}
\email{asingh.iitkgp@gmail.com}
\affiliation{Benjamin Levich Institute, CUNY City College of New York, New York, NY 10031, USA}
\affiliation{Institute for Molecular Engineering, University of Chicago, Chicago, Illinois 60637, USA}
\affiliation{James Franck Institute, University of Chicago, Chicago, Illinois 60637, USA}
\author{Lou Kondic}
\email{kondic@njit.edu}
\affiliation{Department of Mathematical Sciences, New Jersey Institute of Technology, Newark, NJ 07102, USA}
\author{Konstantin Mischaikow}
\email{mischaik@math.rutgers.edu}
\affiliation{Department of Mathematics, Rutgers, The State University of New Jersey, Piscataway, NJ, 08854, USA}
\author{Jeffrey F. Morris}
\email{morris@ccny.cuny.edu}
\affiliation{Benjamin Levich Institute, CUNY City College of New York, New York, NY 10031, USA}
\affiliation{Department of Chemical Engineering, CUNY City College of New York, New York, NY 10031}
\date{\today} 

\begin{abstract} 
Dense, stabilized, frictional particulate suspensions in a viscous liquid undergo increasingly strong continuous shear thickening (CST) as the solid packing fraction, $\phi$, increases above a critical volume fraction, and discontinuous shear thickening (DST) is observed for even
higher packing fractions. Recent studies have related shear thickening to a transition from mostly lubricated to predominantly frictional contacts with the increase in stress. The rheology and networks of frictional forces from two and three dimensional simulations of shear-thickening suspensions are studied. 
These are analyzed using measures of the topology of the network, including tools of persistent homology.
We observe that at low stress the frictional interaction networks are predominantly quasi-linear along the compression axis. With an increase in stress, the force networks become more isotropic, forming loops in addition to chain-like structures.  The
topological measures of Betti numbers and total persistence provide a compact means of describing the mean properties of the frictional force networks, and provide a key link between macroscopic rheology and the microscopic interactions.
A total persistence measure describing the significance of loops in the force network structure, as a function of stress and packing fraction, shows behavior similar to that of relative viscosity, and displays a scaling law near the jamming fraction for both dimensionalities simulated.  
\end{abstract}

\maketitle

\section{Introduction}
\label{sec:into}
In suspensions of solid particles in viscous liquid, an increase in viscosity with increasing shear rate or imposed stress is well-known, and is called shear thickening. 
If the particles are highly concentrated (or ``dense''), such that the conditions approach the maximum flowable solid fraction ($\phi_A$ or $\phi$ in 2- or 3-dimensions, respectively), a very sharp increase in viscosity over a narrow range of shear rate may occur.  This is termed ``discontinuous shear thickening''. 
Recent simulation studies \cite{seto2013discontinuous, mari2014shear, mari2015nonmonotonic, Mari2015discontinuous, Ness2016shear} and phenomenological theory \cite{bashkirtseva2009rheophysics, wyart2014discontinuous, singh2018constitutive} have shown that a rational basis for the strong shear thickening can be found in a breakdown of lubrication films between particles owing to stress levels which are too large for stabilizing colloidal forces (e.g., grafted chains or surface charges, either of which lead to mutual repulsion) to be effective.  This results in a transition to a much stronger tangential resistance to motion, which is modeled as a Coulomb friction.    

These studies find a  relationship between strong shear thickening and shear jamming, in the sense that thickening may be a precursor to jamming: increase of either stress or solid fraction from a shear-thickening condition may lead to shear jamming. To be specific, shear jamming is the phenomenon of generating  a solid structure under shear, but with this solid being fragile in the 
sense that it loses its ability to bear a shearing load (stress) without deformation if  the load direction is changed sufficiently \cite{cates98}.  This loss of solidity is a temporary material failure, as  rearrangement of structure, and in particular of the contact network, should once again result in jamming in the new direction of 
motion~\cite{seto2019shear,singh2019yielding}.   Furthermore, one may wonder whether it is the contact network that is relevant to the process of shear-thickening and shear-induced jamming, or 
whether one needs to consider not only contacts, but also their strength.  If strength of the contacts is considered as well, the concept of a weighted
interaction network is  introduced.
Examples include weak or strong interaction networks considered
commonly when analyzing dry granular matter, with force chains being the most commonly considered structures. 

In shear thickening suspensions, a large change in properties has been associated with formation of a contact network driven by increasing shear stress.  This prompts us to focus on the connectivity properties of this network, 
and its relation to stress transmission.  Essential questions relate to how the connectivity arises, how it depends on the interaction force strength, what are the geometrical and topological properties of the connected structures, and how these structures evolve with both 
the solid fraction and the imposed stress.   All of these issues must be examined for a specific model of the microscopic physics.  

We apply a simulation model that, using a minimal set of
ingredients, has been found to closely approximate behavior observed in experiments  \cite{seto2013discontinuous, mari2014shear, mari2015nonmonotonic, Mari2015discontinuous}.
This model assumes that particle surface separation is maintained by a repulsive force, and is thus lubricated by the liquid film that remains, until the applied stress overwhelms this repulsive force \cite{seto2013discontinuous,mari2014shear}. In practice, many suspensions have steric or electrostatic forces, repulsive in either case, between particles to stabilize against surface contact, reducing aggregation or contact interactions. This leads to better flowability, i.e., less tendency to jam and lower viscosity. Such a repulsive force, of magnitude $F_0$, gives rise to a stress scale $\sigma_0=F_0/6\pi a^2$ for particles of radius $a$, and within our simulational model or the theory of Wyart \& Cates \cite{wyart2014discontinuous}, this stress level marks the 
crossover from lubricated to contact frictional interactions  between particle surfaces. At low stress, $\sigma \ll \sigma_0$, particle interactions are lubricated (and hence frictionless) and the viscosity is relatively low, diverging at random close packing (RCP) at a solid fraction that we denote by $\phi_{\rm J}^0$, with the subscript J indicating jamming. On the other hand, at high stress, $\sigma \gg \sigma_0$, most close interactions are frictional contacts, and the viscosity
is both much larger and diverges at a solid fraction $\phi_{\rm J}^\mu < \phi_{\rm J}^0$; $\phi_{\rm J}^0$ depends on the interparticle friction~\cite{mari2014shear}. With increase in stress, the transition  between these
 two states results in shear thickening. Based on the value of packing fraction $\phi$ relative to the frictional jamming point $\phi_{\rm J}^\mu$, different forms  of shear--thickening can be observed.
 For $\phi \ll \phi_{\rm J}^\mu$, the relation between shear rate $\dot{\gamma}$ and shear stress $\sigma$ is monotonic, leading to continuous shear thickening (CST), while for larger 
 $\phi$ the nore abrupt discontinuous shear thickening (DST) may occur.

To gain better insight to the lubricated-to-frictional rheology transition, we apply tools of topology, with a focus on the network theoretical methodology of {\em persistent homology}, which has previously been applied to the closely related 
question of jamming in dry granular systems~\cite{physicaD14,pre13,pre14}.   Persistent homology provides precisely defined and quantitative 
measures of the global interaction network, which is built on top of the contact network,  as discussed in Section~\ref{subsec:persistent}.   
 In the present context, we use these tools to characterize flowing steady states, and thus the 
statistical properties of the interaction networks formed under different conditions in shear-thickening suspensions.   

We point out two features of the analysis of interaction networks based on persistent homology that 
distinguish it from alternative approaches.  First, this type of analysis naturally includes information about 
the forces between particles, and allows for quantification and comparison of the interaction 
networks, both between different states of a system, and between different systems.  Second, 
this approach applies to both two- and three-dimensional ($2$D and $3$D) phenomena.  This is convenient, since 
the $2$D case allows for development of intuition and visual demonstration of the tools applied here, while 
$3$D results are in accord with physical experiment.  The ability of the persistent homology tools to be applied without change of definition to $3$D contact networks is an advantage for this method over complementary approaches 
to exploration of networks;  see~\cite{networks_review_18} for a review.    We develop this approach here, demonstrating that measures of 
the structure determined by persistent homology show sharp changes where the rheology of the material undergoes abrupt shear thickening.   

\section{Background \& methods}
\label{sec:methods}
\subsection{Suspension flow simulation}

We simulate simple shear flow of an assembly of non-Brownian frictional spheres (a monolayer in the two dimensional case) immersed in a Newtonian fluid under an imposed stress using Lees-Edwards periodic boundary conditions \cite{lees1972computer}.
To avoid ordering, bidisperse  particles with radii $a$ and $1.4a$ are mixed at equal volume fractions.  The particles interact through near-field hydrodynamic forces (i.e., lubrication forces) and frictional contacts.
This simulation scheme has been shown to reproduce many features seen  experimentally in shear-thickening of dense suspensions  \cite{seto2013discontinuous, mari2014shear}.
We consider the motion to be inertialess (at zero Reynolds number) so that the equation of motion reduces to force balance between hydrodynamic ($\mathbf {F}_{\rm H}$) 
and contact ($\mathbf {F}_{\rm C}$) forces on each particle,
\begin{equation}
 0 = \mathbf {F}_{\rm H} (\mathbf{r,u}) + \mathbf {F}_{\rm C} (\mathbf{r}) ~,
\label{eq-1}
\end{equation}
where $\mathbf{u}$ and ${\mathbf{r}}$ are written here as the many-body position and velocity vectors ($\mathbf{u} \equiv \dot{\mathbf{r}}$). A similar torque balance applies.
To allow contact, the lubrication resistance singularity is cut off at a surface
 separation $h=10^{-3}a$. The contact  interactions are modeled using the approach of Cundall \& Strack \cite{cundall1979discrete}. 
 The tangential force between two particles is limited according to the Coulomb friction law to be no larger than the friction coefficient $\mu$ times the normal force, i.e. $\bigl| \mathbf{F}_{\mathrm{C}}^{t} \bigr| \leq \mu \bigl| \mathbf{F}_{\mathrm{C}}^{n}\bigr|$, for compressive normal forces.
 To incorporate rate dependence, we employ a Critical Load Model (CLM), in which an interparticle normal force threshold $F_{0}$ must be applied to activate friction between particles \cite{seto2013discontinuous, mari2014shear, Ness2016shear}. In this study we use an interparticle friction coefficient of $\mu=1.0$. 
The apparent viscosity of the suspension is defined as $\eta = \sigma/\dot{\gamma}$, where $\dot{\gamma}$ is the shear rate of the simple shear flow, i.e., $d u_x/dy$. 
 The viscosity is presented through its form relative to the fluid viscosity $\eta_r = \eta/\eta_0$, where $\eta_0$ is the pure fluid value. A large increase in $\eta_r$ with increasing shear rate is found, and this is the shear thickening of interest. 
 \par
 The force $F_0$ sets a stress scale, $\sigma_0 = F_{\rm 0}/6\pi a^2$, 
 for the onset of shear thickening. In the remainder of this work, we report quantities in terms of   $\tilde{\sigma} = \sigma/\sigma_0$ and the scaled strain rate $\dot{\gamma}/\dot{\gamma}_0$ with $\dot{\gamma}_0 \equiv F_{\rm 0}/{6\pi \eta_0 a^2}$. 
 We normalize the force by the imposed shear stress $F = F/\tilde{\sigma}a^2$.
 We use $N = 2000$ particles for 2D simulations and $N = 500$  particles for 3D simulations in a unit cell; to test size dependence, $N = 1000$ and 2000 were considered in 3D for a few cases.

\begin{figure*}
\centering
 \subfigure[]{
 \includegraphics[width=.45\textwidth]{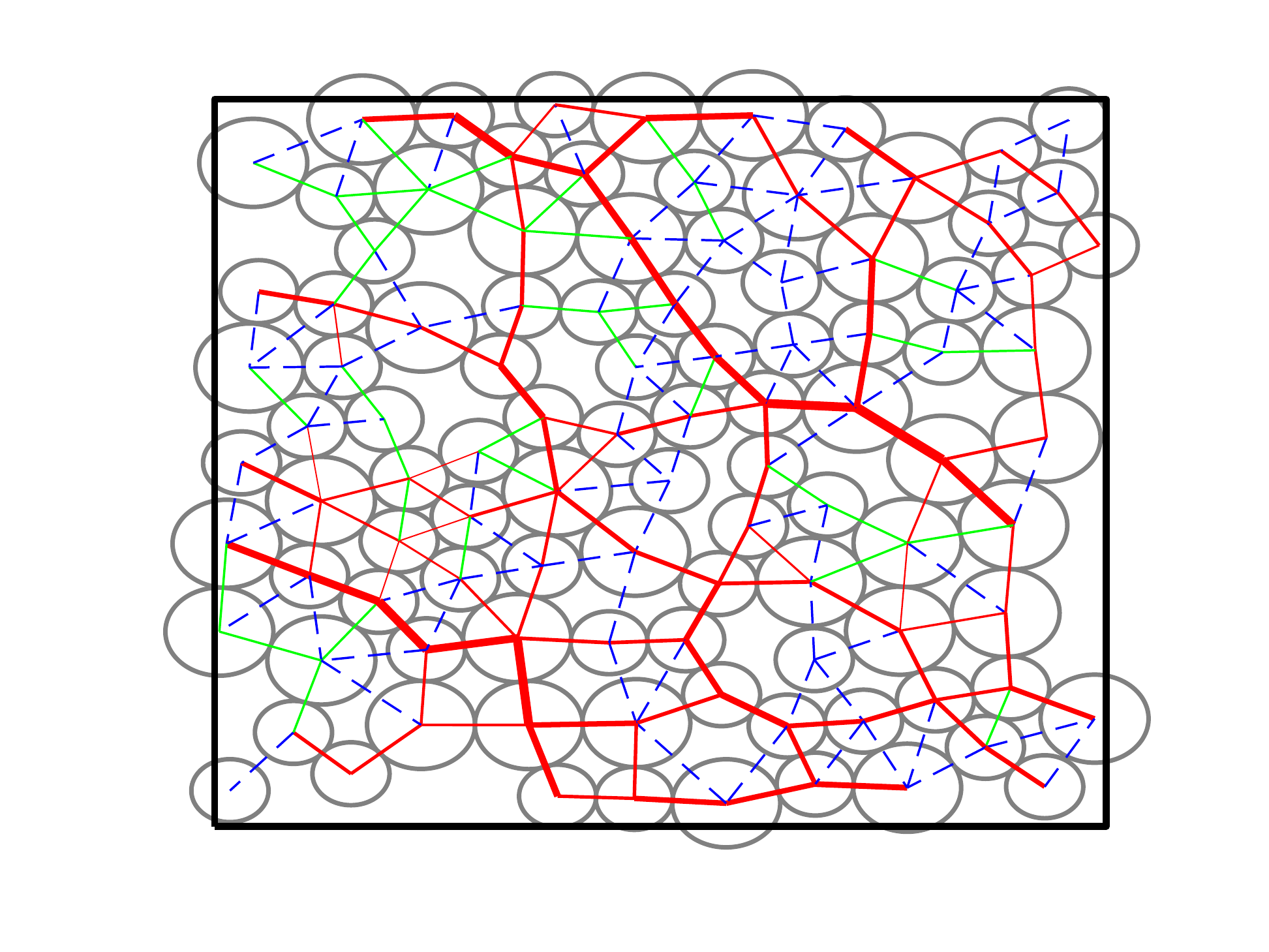}}
 \subfigure[]{
 \includegraphics[width=.45\textwidth]{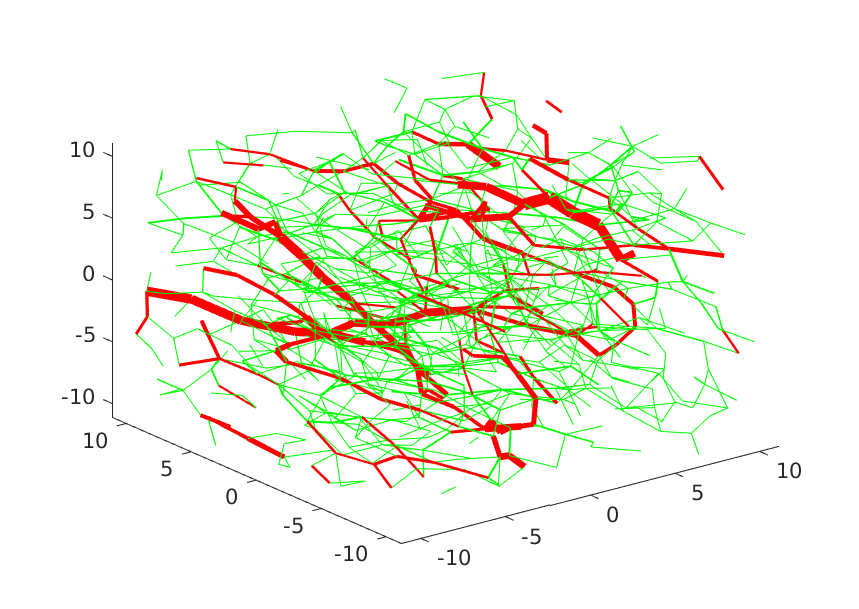}}
\subfigure[]{
\includegraphics[width=.45\textwidth]{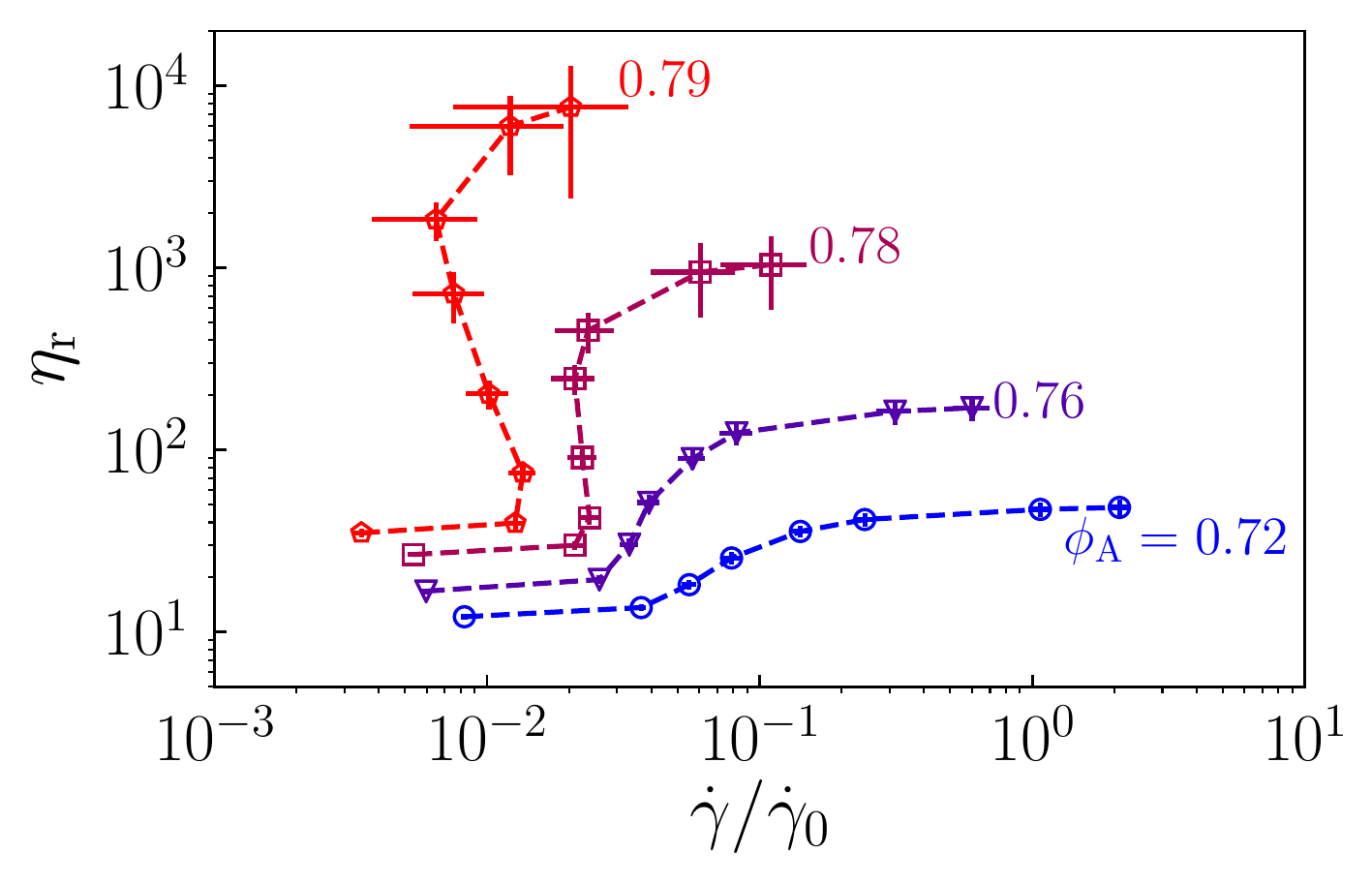}}
\subfigure[]{
\includegraphics[width=.45\textwidth]{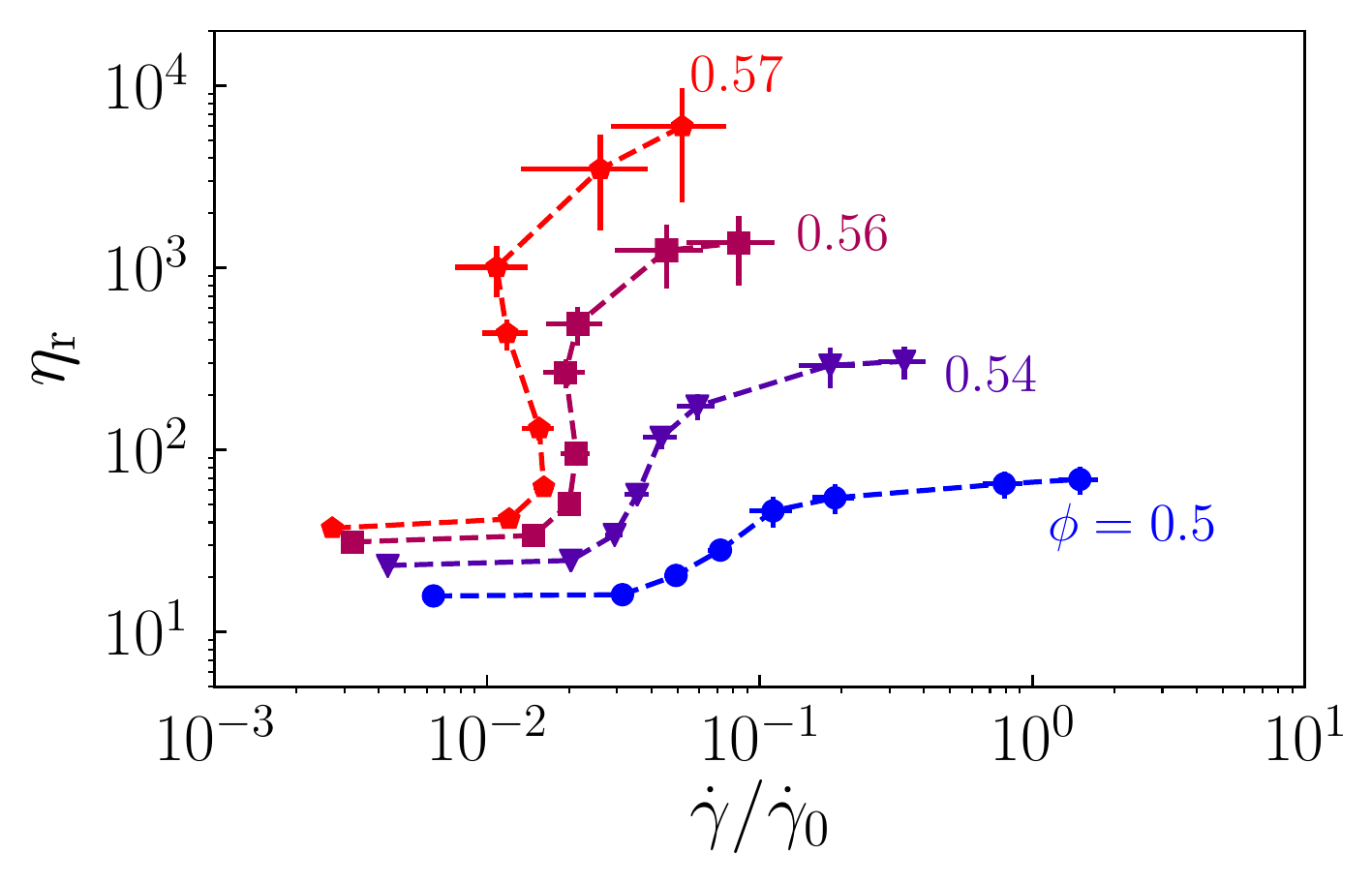}}
\caption{
(a-b) Snapshots of the force  network at $\tilde{\sigma} =1$ for (a) $\phi_{\rm A} = 0.78$ in 2D and (b) $\phi =0.56$ in 3D. Line thickness is proportional to the normalized force $F$ with red, green and blue lines showing frictional, frictionless contact, and lubrication interactions, respectively.
(c-d) Relative viscosity, $\eta_{\rm r}$, as a function of dimensionless shear rate 
$\dot{\gamma}/\dot{\gamma}_0$ in stress-controlled simulations in (c) 2D and (d) 3D for several packing fractions.
}
\label{fig-1}
\end{figure*}

 From the simulations, we determine particle positions, normal and  tangential contact forces, and non-contact lubrication forces. 
 We perform simulation at a given condition for 20 strain units and record the data at strain steps of $0.01$.
  Figures~\ref{fig-1}(a) and~\ref{fig-1}(b) show snapshots from simulation with
 $\tilde{\sigma} =1$ for packing fractions $\phi_{\rm A} = 0.78$ and $\phi = 0.56$ in 2D and 3D, respectively.
 In these images, we show the frictional (red solid line) and frictionless contacts (dashed  green lines), and lubrication interactions (blue dashed lines).
  The thickness of each frictional segment is proportional to the normalized force level 
 $F$ acting at the contact.  As we will see, 
 the development of the contact network with increasing stress is responsible for significant changes in the rheological response of the suspension.  
 
In the absence of any non-hydrodynamic force, the suspension would be expected to exhibit a rate- or stress-independent relative viscosity, with close interactions lubricated. 
Similarly, hydrodynamic interactions together with  frictional contacts, but without the repulsive force, result in a rate-independent rheology. 
The rheological rate-dependence thus arises due to the shear force overwhelming the repulsive forces, modeled here by the CLM approach \cite{mari2014shear}, to cause a change in the dominant stress generation mechanism.  In our simulations, we use a controlled shear stress  protocol \cite{mari2015nonmonotonic}, in which the shear rate $\dot{\gamma}$ fluctuates. The relative viscosity is given by $\eta_r = \tilde{\sigma}/\langle \dot{\gamma}/\dot{\gamma}_0\rangle$, where angle brackets imply time average. 

  In order to present the results of topology-based analysis in a visually transparent manner, we  consider 2D before turning to 3D simulations, for which comparisons of the rheology to experimental work have been made \cite{mari2014shear,Mari2015discontinuous}. 
 Figures~\ref{fig-1}(c) and~\ref{fig-1}(d) show $\eta_r$ as a function of shear rate for four packing fractions in 2D and 3D, respectively.
For the lowest packing fractions,  namely $\phi_{\rm A} \le 0.76$ in 2D and $\phi \le 0.54$ in 3D, $\eta_r$ increases continuously between two rate-independent values: this is CST.  At packing fractions of $\phi_{\rm A} = 0.78$ in 2D and $\phi = 0.56$ in 3D, the viscosity increases at almost
 constant shear rate over a brief interval, indicating that these are approximately the packing
fractions for onset of DST. For higher packing fractions, $\phi_{\rm A} = 0.79$ and $\phi = 0.57$, the viscosity function becomes sigmoidal under  imposed stress, indicating that the suspension is fully in the DST state~\citep{singh2018constitutive}. Note that if the simulations at the larger fractions are performed at fixed rate,  $\eta_r(\dot{\gamma}/\dot{\gamma}_0$) is observed to change discontinuously rather than following the S-shaped curve~\citep{mari2015nonmonotonic, wyart2014discontinuous, singh2018constitutive}.
 
\begin{figure*}
 \centering
 \includegraphics[width=.6\textwidth]{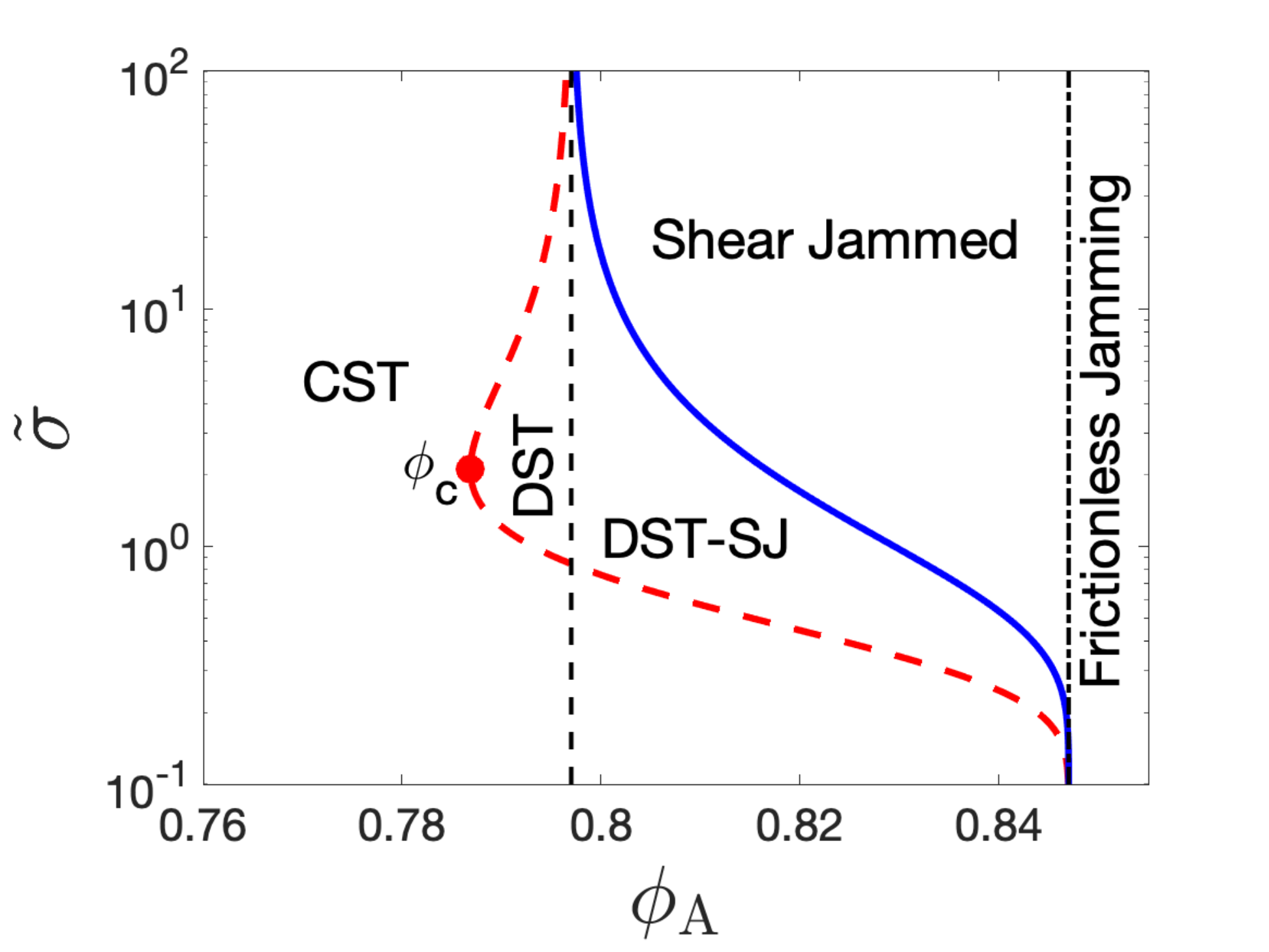}
 \caption{Flow-state diagram in the shear stress versus 2D packing fraction
 $(\tilde{\sigma},\phi_{\rm A})$ plane for $\mu=1$; for the 3D version, see \cite{singh2018constitutive}. The red dashed line shows the locus of points for DST (where $\partial \dot\gamma/\partial \sigma= 0$) and the leftmost point on this curve is the minimum packing fraction, $\phi_{\rm C}$, at which DST is observed.
 The blue solid curve shows packing fraction dependent 
 maximum flowable stress above which the suspension is shear-jammed. Dashed and dotted-dashed black lines represent frictional $(\phi_{\rm J}^\mu=0.795)$ and frictionless $(\phi_{\rm J}^0=0.855)$ jamming points in 2D, respectively.}
 \label{fig-2}       
 \end{figure*}

 Figure~\ref{fig-2} shows a flow-state diagram, based on simulation results and a recently proposed model~\cite{singh2018constitutive}; this particular diagram is for 2D, with the 3D version in the  noted reference showing the same features.  
 As discussed above, for the lower packing fractions $\phi < \phi_{\rm C}$, CST is observed. For packing fractions $ \phi_{\rm C} \le \phi < \phi_{\rm J}^\mu$, DST is observed between two flowing states. For this range of $\phi$,
a curve shows the locus of DST points, i.e., ${d\dot\gamma/d\sigma}= 0$. 
For $\phi > \phi_{\rm J}^\mu$, DST is observed between a flowing state (at low stress) and a shear jammed state and is termed DST--SJ.
The stress required to observe DST as well as shear jamming decreases
 with an increase in packing fraction. Both stresses vanish upon the approach to the frictionless jamming  point.  We do not consider the 
 DST--SJ 
 state further in this work, as the growth of the contact network with increasing stress is conceptually similar to the ``pure'' DST regime, 
 while the subtleties of the network structure at jamming would expand beyond our desired scope.  The goal here is to demonstrate the utility of 
topological data analysis as a tool for describing underlying features of the interaction networks resulting in the rheology of shear-thickening suspensions.

 \begin{figure*}
 \centering
  \includegraphics[width=.94\textwidth]{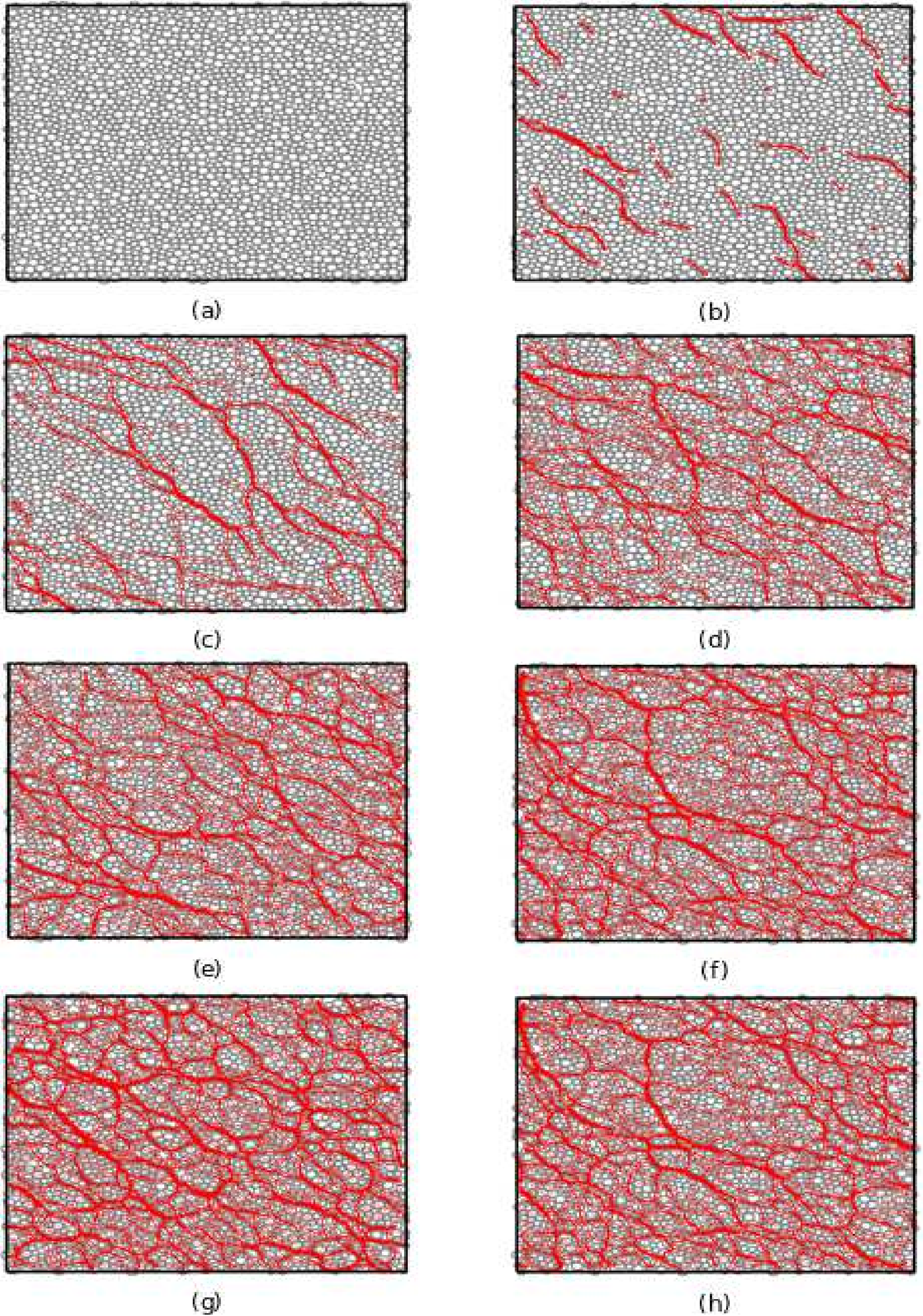}
 \caption{Interaction network of all frictional contacts (normal force above the critical value, $F_0$) corresponding to a snapshot in steady state obtained in simulations with $\tilde{\sigma} =$ (a) 0.1, (b) 0.5,
 (c)1.0, (d) 2.0, (e) 5.0, (f) 10, (g) 50, and (h) 100 in 2D. Line thickness is proportional to the normalized force, 
 $F$.}
 \label{fig-3}       
 \end{figure*}

Recent studies have linked shear thickening with a growing number of frictional contacts as shear 
stress increases \cite{seto2013discontinuous, mari2014shear, wyart2014discontinuous,Ness2016shear}. Only limited work has explored the network properties of sheared suspensions \cite{Edens_2019,thomas2018microscopic}. In the following, we 
focus on evolution of the network structure formed by the frictional forces.  These are readily visualized in 
$2$D, and Figure~\ref{fig-3} shows how the contact interaction network builds in 2D for stresses varying through the DST transition at $\phi_{\rm A} = 0.79$.   
At low stress $\tilde{\sigma}=0.1$,  frictional contacts
are absent, so that the suspension is in the low viscosity frictionless state. For $\tilde{\sigma}=0.5$ and 1.0, the suspension begins to shear thicken, and frictional forces 
appear as roughly linear structures (force chains) along the compression axis, i.e. along $y = -x$  of the simple shear flow $u_x = \dot{\gamma}y$.
With increase in stress to $\tilde{\sigma}\ge 2.0$, the suspension enters the DST regime and frictional interaction networks begin to form loops.
With further increase in stress, the number of loops increases and saturates for $\tilde{\sigma}>10$. Combined with the rheological 
curves in Fig.~\ref{fig-1}c, this illustrates the correlation between strong shear thickening and frictional interaction networks.  Note that all structures described are transient, continuously forming, flowing, and breaking as the result of the bulk shearing motion.

\subsection{Overview of topological measures: From Betti numbers to persistent homology}
\label{subsec:persistent}

\begin{figure*}
\centering
\subfigure[]{
\includegraphics[width=.37\textwidth]{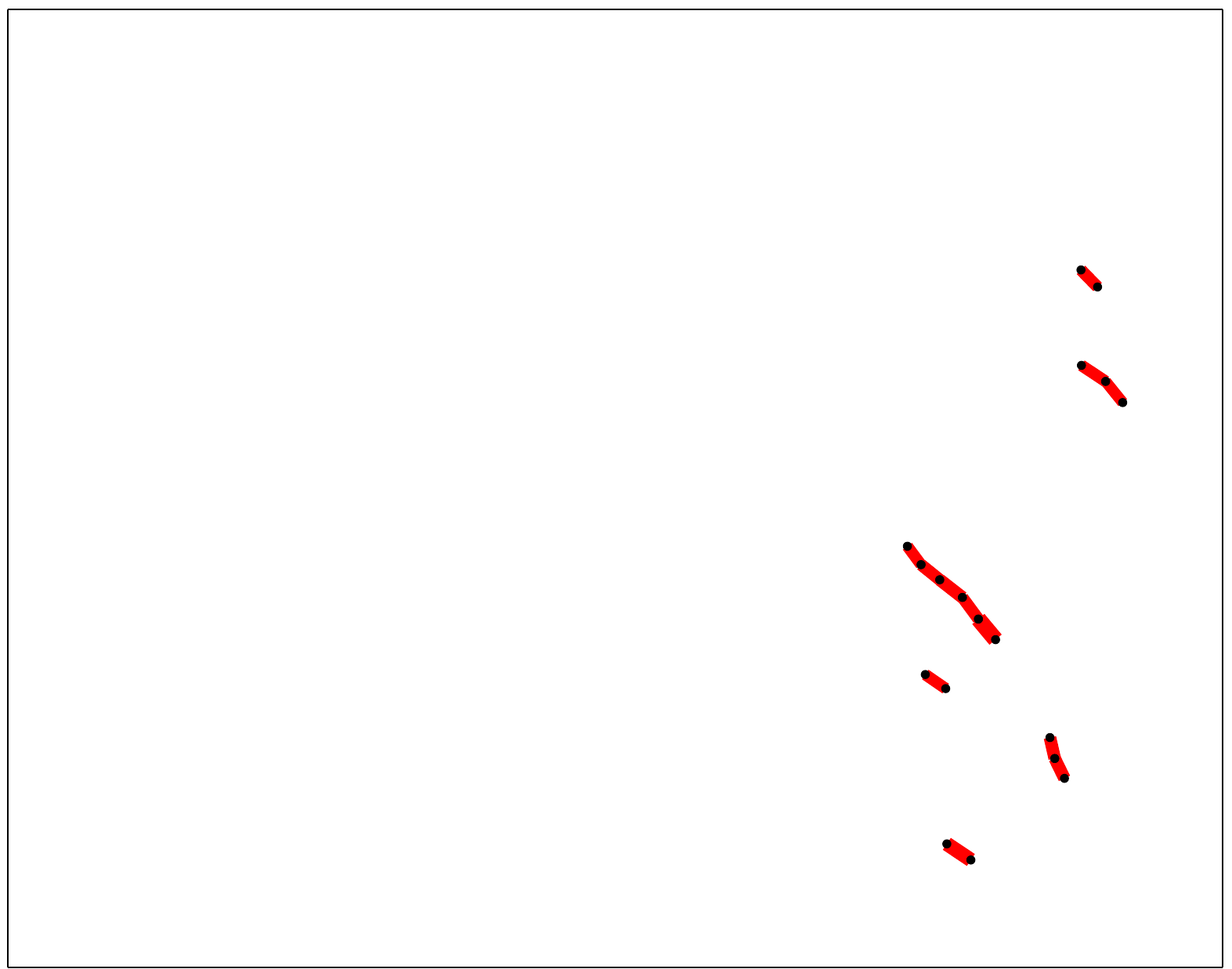}}
\subfigure[]{
\includegraphics[width=.37\textwidth]{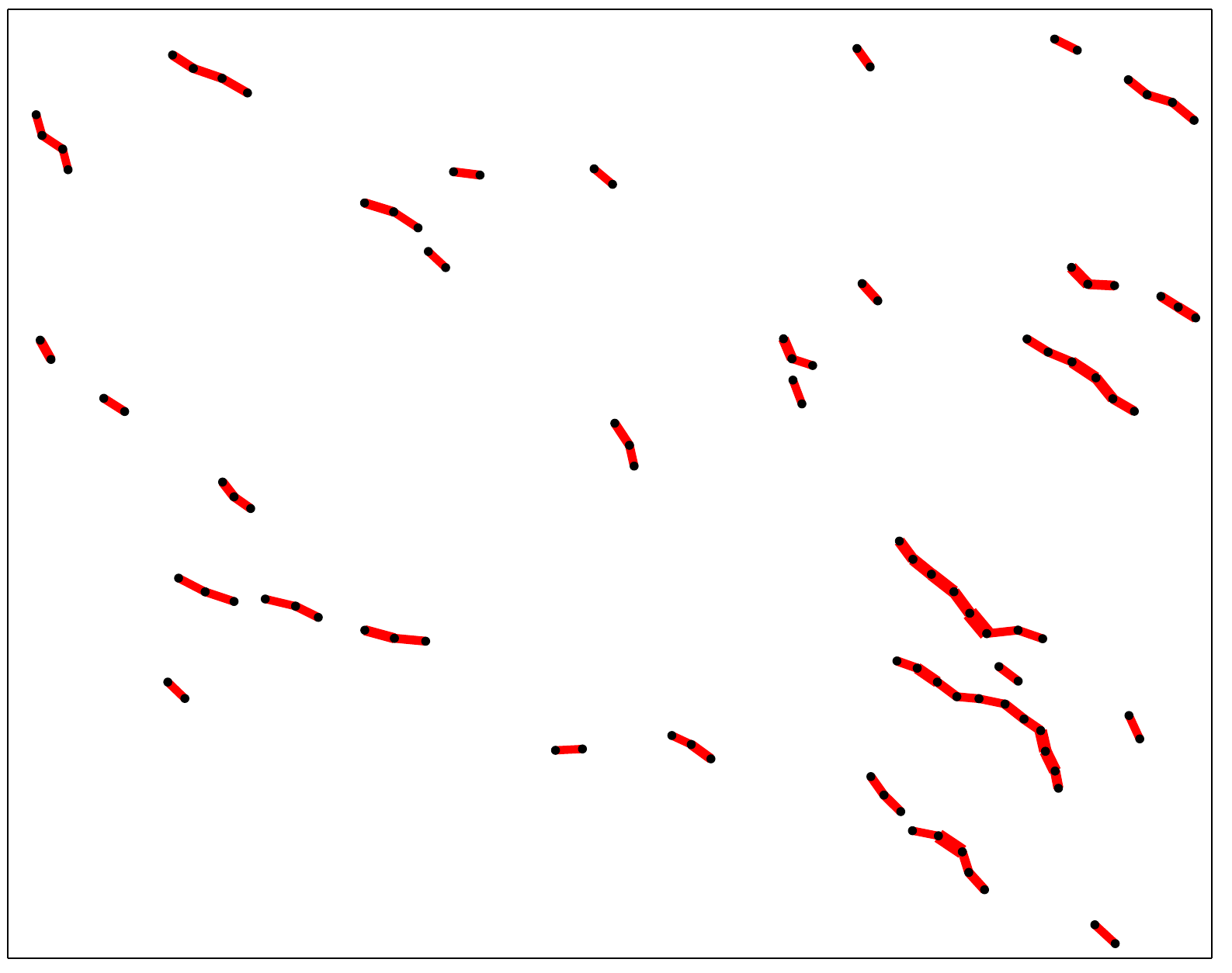}}
\subfigure[]{
\includegraphics[width=.37\textwidth]{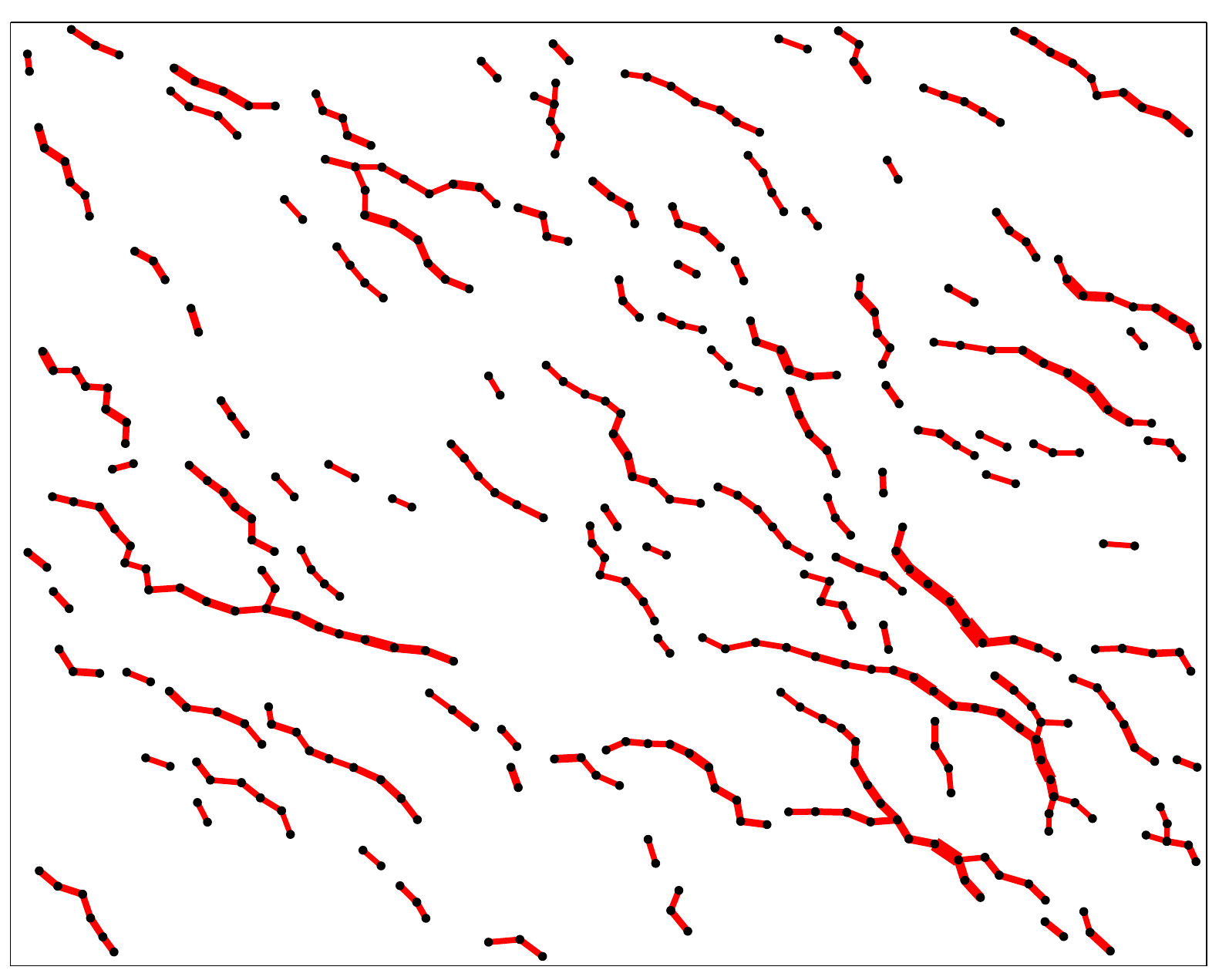}}
\subfigure[]{
\includegraphics[width=.37\textwidth]{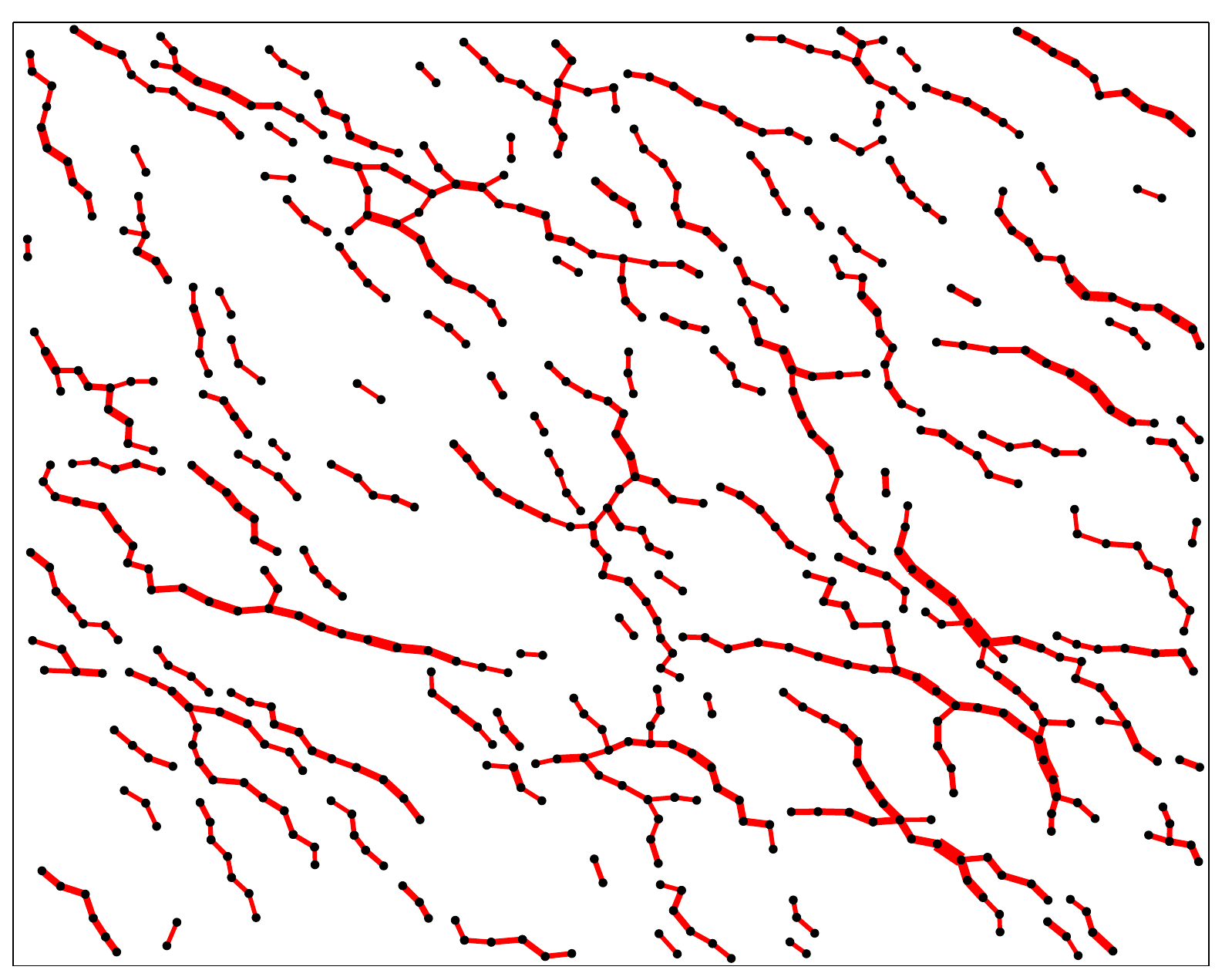}}
\subfigure[]{
\includegraphics[width=.37\textwidth]{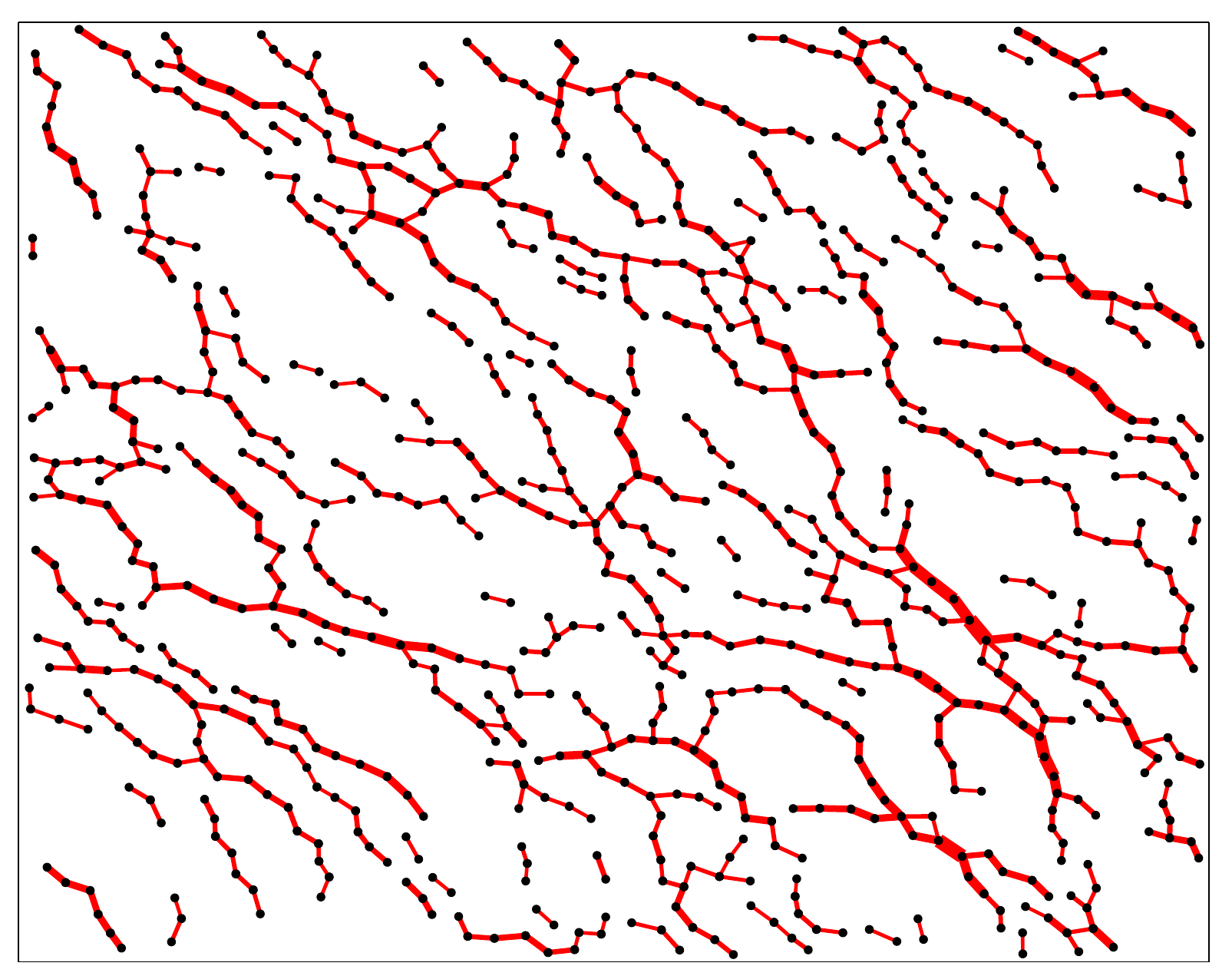}}
\subfigure[]{
\includegraphics[width=.37\textwidth]{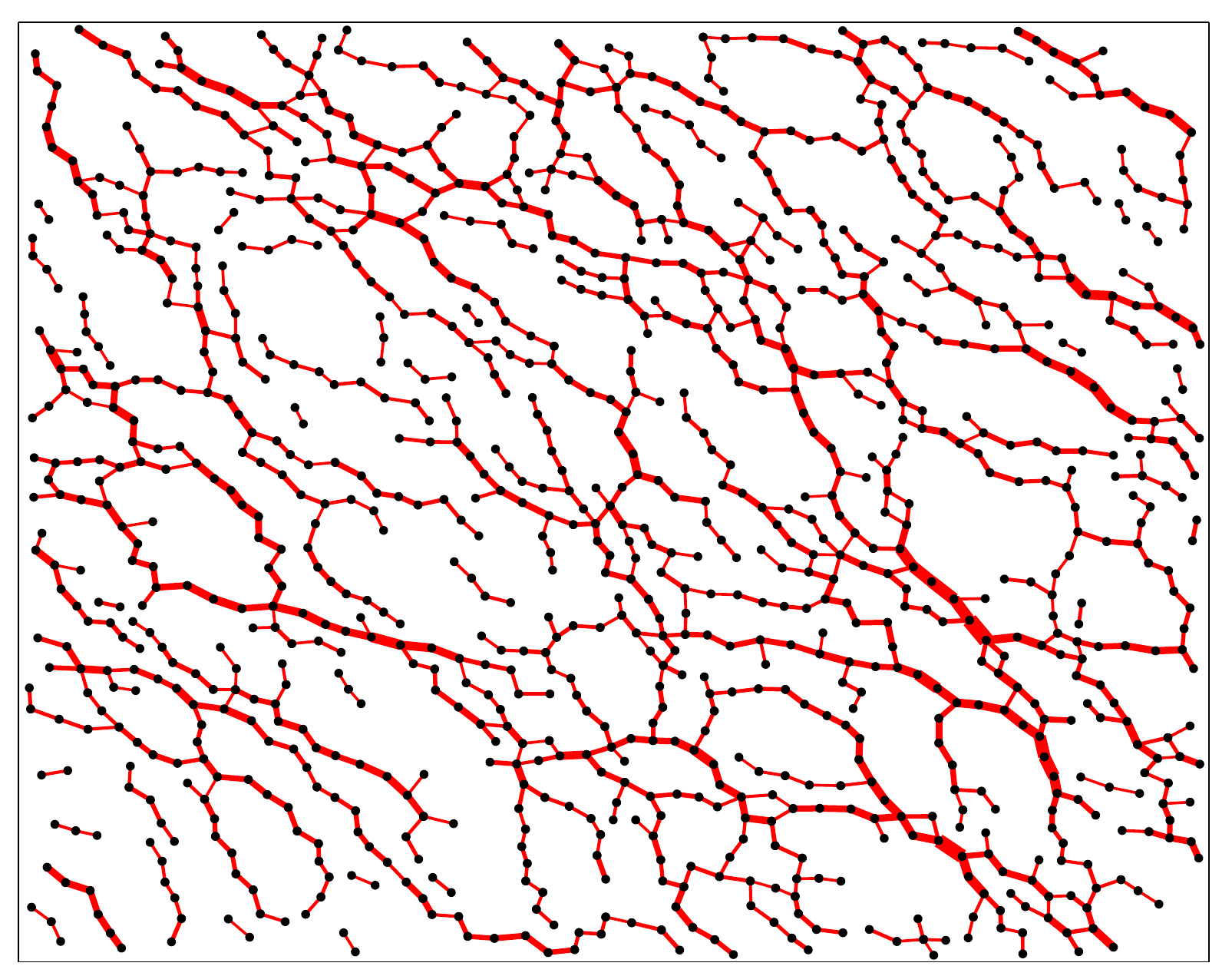}}
\subfigure[]{
\includegraphics[width=.37\textwidth]{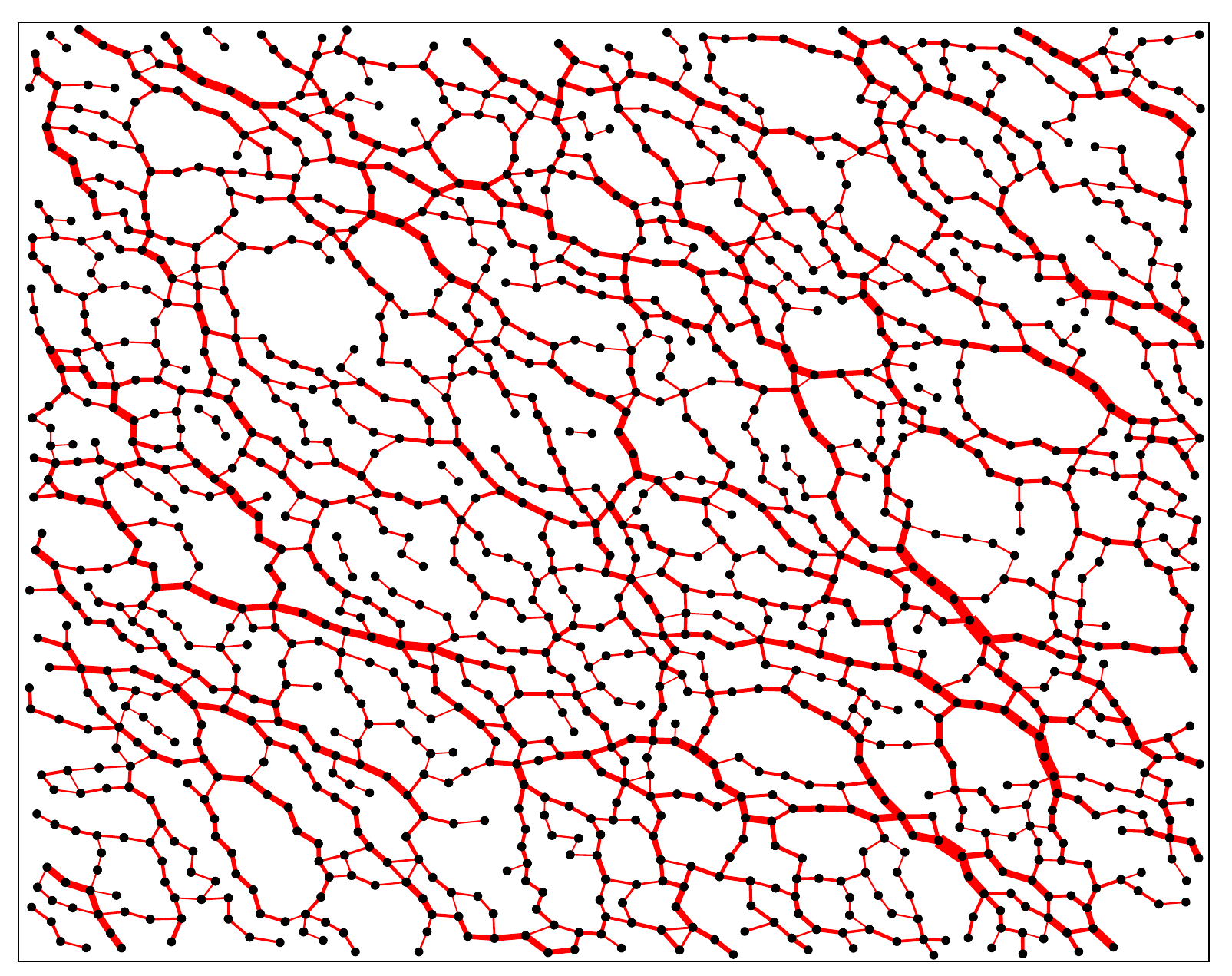}}
\subfigure[]{
\includegraphics[width=.37\textwidth]{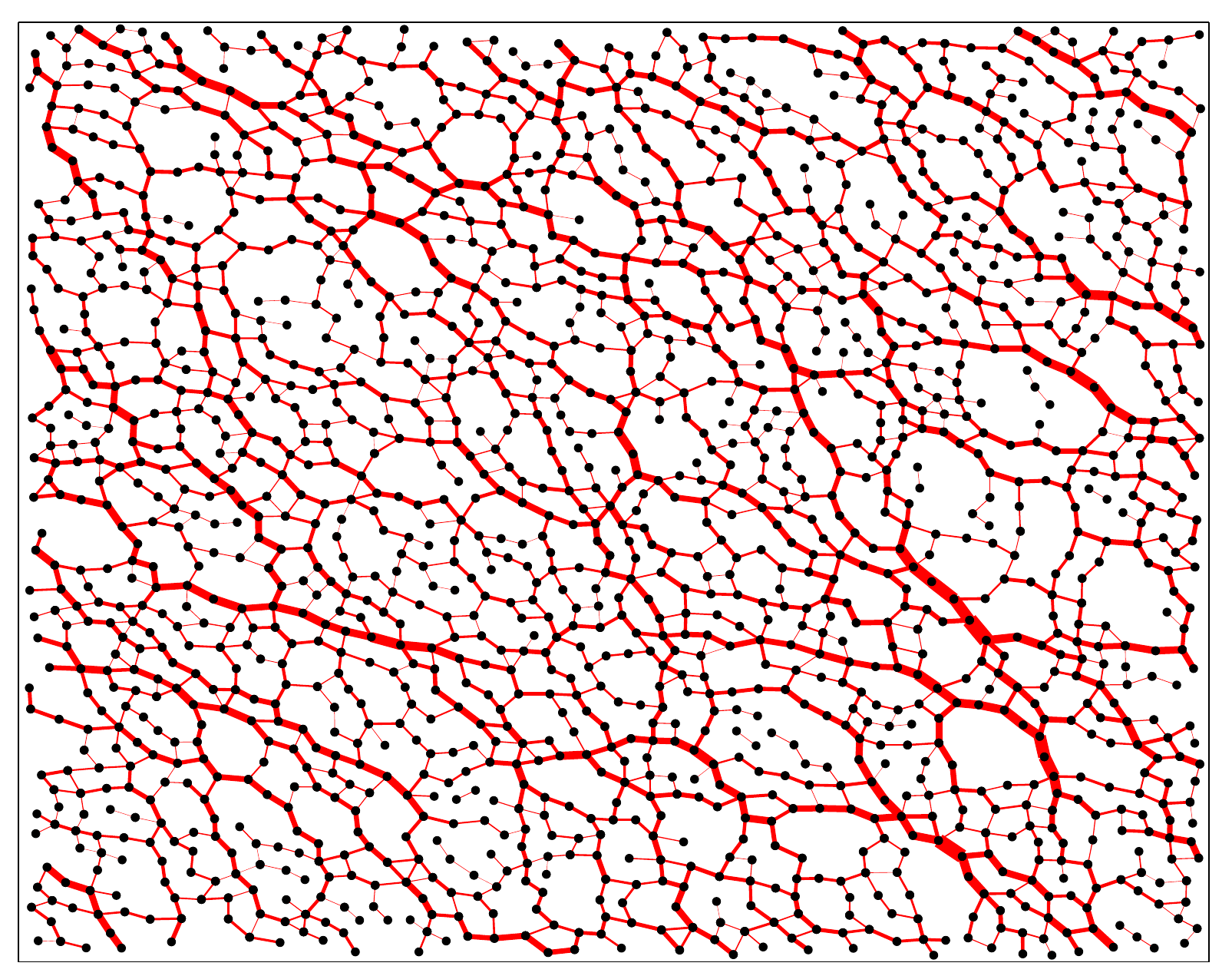}}
\vspace{-0.15in}
\caption{Filtration of the interaction network of all frictional contacts 
for $\phi=\phi_{\rm A}= 0.76$ and $\tilde{\sigma}=10$ corresponding to a snapshot in steady state, for several values of the threshold, in order left to right and top to bottom: $F=$ (a) $4$, (b) $3$, (c) $2$, (d) $1.6$, (e) $1.3$, (f) $1$, (g) $0.5$, and (h) $0$.  Here, black dots show particle centers and thickness of a red line corresponds to the strength of the
normal force at the contact.
}
\label{fig:2d_filtration}
\end{figure*}

To introduce the relevant concepts, consider Figure~\ref{fig:2d_filtration}, which shows the interaction networks as a function of the interaction level, an {\em interaction network filtration} or simply a {\em filtration}; here the interparticle force is scaled by imposed stress $\tilde{\sigma}a^2$.   This figure illustrates the 
portions of the contact network experiencing force larger than the associated threshold or filtration level.  At a very large force, only the 
most highly loaded contacts and the particles they connect are shown,  and these are isolated objects.  As the threshold is reduced, these 
components `grow' and eventually merge with others.   The information contained in such figures for a set of filtration levels 
will be used in what follows to formulate the relevant topological measures.  
Note that there is a visual similarity between the filtration at decreasing force threshold in Figure  \ref{fig:2d_filtration} with the contact networks formed at increasing stress level in Figure \ref{fig-3}: although  structural change takes place with increasing stress, the implication of this similarity  is that the type of structures which are able to achieve the necessary force to generate a contact at low stress generate  contacts which have the largest forces at elevated 
stresses where a fully ramified network forms.

The numbers of structures of a particular type can be described by Betti numbers.   
The number of connected structures of any sort, from a contacting pair to an extended branched chain, is given by the zeroth Betti number, $\beta_0$, while the number of 
structures which form a closed loop is given by the first Betti number, $\beta_1$.
The structures formed by connected particles enclosing a volume is given by the second Betti number, $\beta_2$, which is relevant 
in $3$D, but not in 2D.  We find no such structures in our 3D simulations ({\em i.e.}, $\beta_2 =0$), and thus we only consider $\beta_0$ and $\beta_1$ in the following.  
\par 

The manner of 
defining the structures is associated with a decreasing force threshold.  As the force threshold $F$ decreases, `birth' of a connected 
structure of a given Betti number is associated with the force threshold in a filtration for which the structure appears, i.e., it is the largest 
force within the structure.  As the threshold is further decreased, the structure may grow in size as additional contacts are added.  
Consider a $\beta_0$ structure, which formed at $F_1$, and a second which formed at $F_2<F_1$; if these two structures merge, the 
resulting union results in the `death' of one structure, and by definition it is the younger structure, which formed `later' in the filtration (at lower force in a sequence of progressively reduced force threshold) which 
is lost, i.e., the structure with birth at $F_2$ dies upon the merging.  The persistence of structures through the reduction of force threshold 
gives rise to the term persistent homology.  

\begin{figure*}
\centering
\subfigure[]{
\includegraphics[width=.45\textwidth]{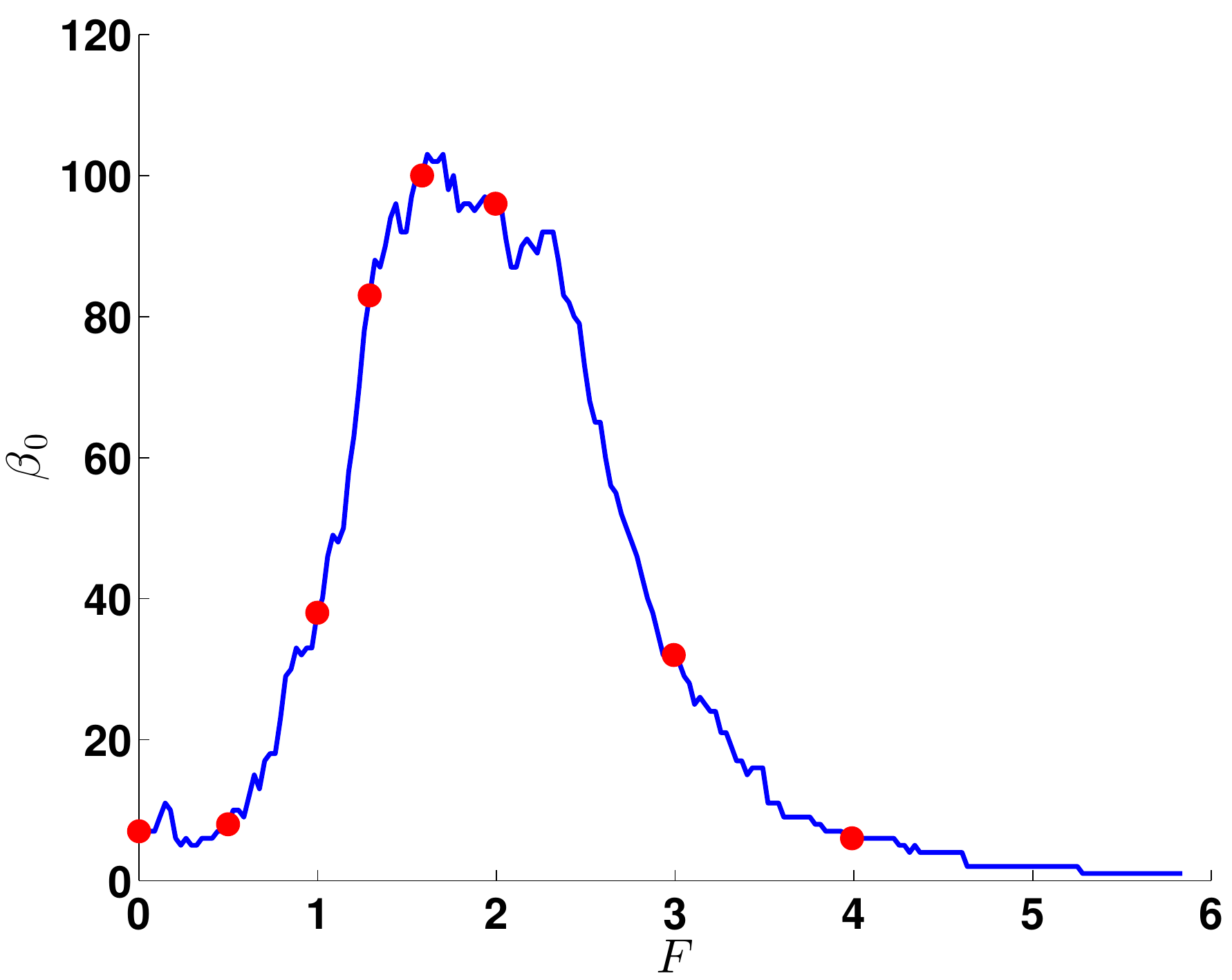}}
\subfigure[]{
\includegraphics[width=.45\textwidth]{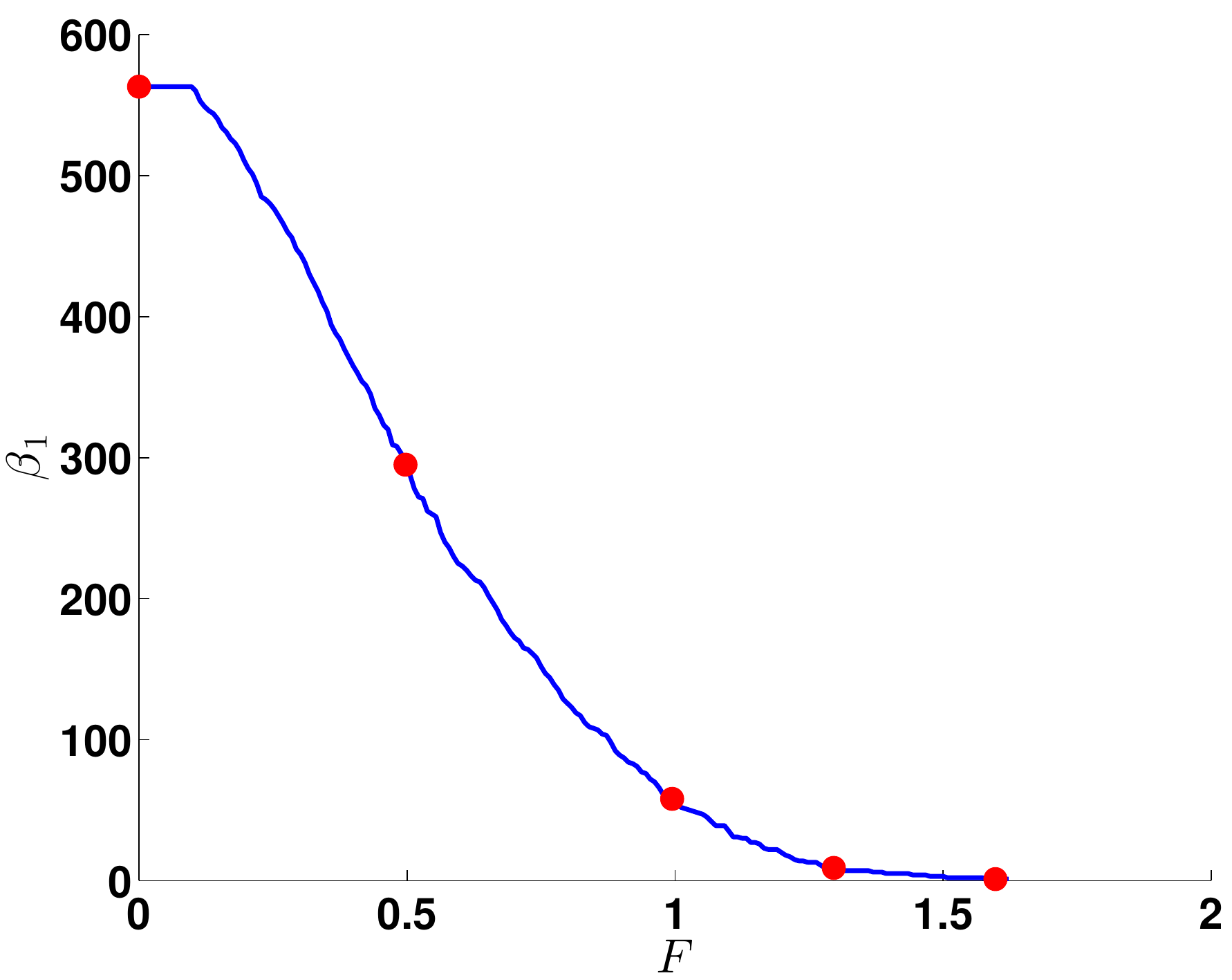}}
\caption{Betti numbers corresponding to the filtration in Figure~\ref{fig:2d_filtration} for components (a) and loops (b). 
The (red) dots are at the force threshold levels corresponding to the values illustrated in   Figure~\ref{fig:2d_filtration}.}
\label{fig:betti_numbers}
\end{figure*}

Figure \ref{fig:betti_numbers} shows the Betti numbers for the configuration and network filtrations of Fig.~\ref{fig:2d_filtration}; 
the red points in Fig.~\ref{fig:betti_numbers} correspond to the threshold levels shown 
in Fig.~\ref{fig:2d_filtration}.  
We see from the $\beta_0$ plot that the maximum contact force level is slightly below $F=6$.  As the force threshold is decreased,  $\beta_0$ increases monotonically until $F \approx 2.5$.
Thus we see that connected structures are formed by contacts carrying different levels of force.
The fact that there is a rapid increase in $\beta_0$ until $F \approx 2.2$ indicates that there are many such components, with considerable variation in the force level.

Figure~\ref{fig:betti_numbers}(b) shows that loops are absent until $F\approx 1.5$ and appear in significant numbers only for $F < 1$, which is well below the value at which the number of components starts decreasing rapidly.
This implies that there is significant force heterogeneity, even on the level of single loops.
To understand why, consider the formation of a loop at a given force level.
Typically the loop forms as a single edge is added to a connected structure consisting of edges with higher forces. 
From Fig.~\ref{fig:betti_numbers}(a) and ~\ref{fig:betti_numbers}(b) most loops form after many components of varying force intensity have merged.
Thus, these loops are made up of components with a wide range of force levels.
Furthermore, the fact that $\beta_1$ grows to much larger values than $\beta_0$ shows that the larger connected structures must contain numerous loops.  
This may be thought of as similar to the meshwork structure of a net, although it lacks the regularity of such a structure as shown by Fig.~\ref{fig:2d_filtration}; a single 
connected piece of the mesh may contain from no loops to many loops, and the number and size of the loops is clearly mechanically relevant.

\begin{figure*}
\centering
\subfigure[]{
\includegraphics[width=.45\textwidth]{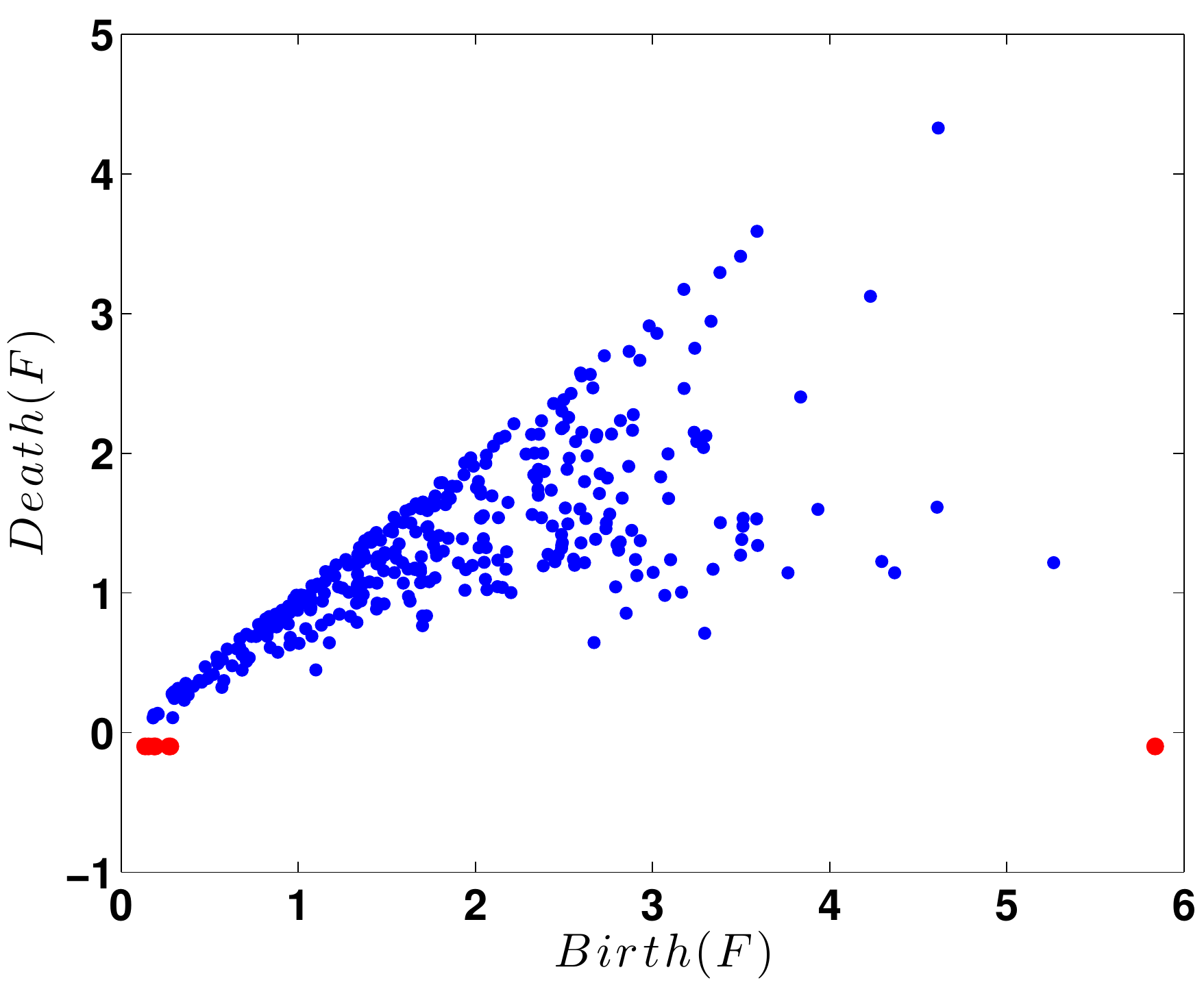}}
\subfigure[]{
\includegraphics[width=.45\textwidth]{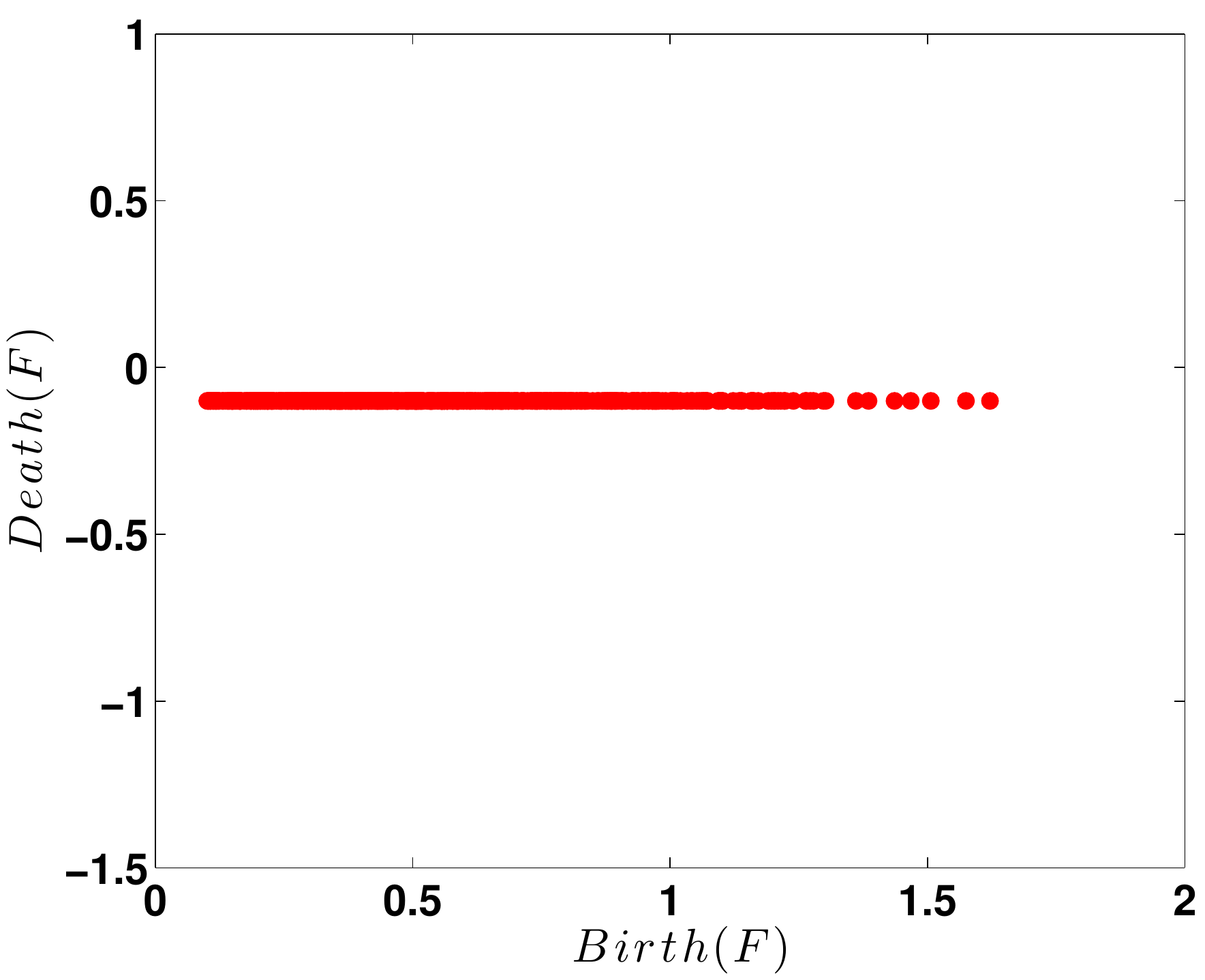}}
\caption{Persistence diagrams corresponding to the filtration in Figure~\ref{fig:2d_filtration} for components (a) and loops (b).}
\label{fig:pers_diags_2d}
\end{figure*}

While Betti numbers provide useful information regarding the number of components and loops, their
utility is somewhat limited by the fact that they depend on the chosen force threshold, and therefore one
needs to consider a large number of force threshold values to reach a reasonably complete understanding of the system. 
In addition, Betti numbers provide only the number of components and loops; they 
do not provide information about how these components and loops are connected.  In principle, the system could
even change its topological properties without a change of Betti numbers, e.g., a component could be born 
at essentially the same force level at which another component dies, and therefore topology changes but $\beta_0$ remains
the same.  
Persistence diagrams allow us to deal with these concerns.

Figure \ref{fig:pers_diags_2d} shows an example of such diagrams for $\beta_0$ and $\beta_1$.    
Here, for each structure, a point is placed at the (birth, death) coordinates $(b,d)$.  If a structure is still present at 
the minimum force threshold of the filtration, it is plotted (in red) with the coordinate pair $(b,-0.1)$ to illustrate that it was born at the threshold level $b$, and  that it never dies.  In Fig.~\ref{fig:pers_diags_2d}(a) the red point with $b \approx 6$ shows the component that formed first (at the largest threshold value), and this component by definition never dies.  The red dots with $b\approx 0$ are specific to the system considered, and they show existence of separate components that form at very low threshold values.  Such structures are possible due to  forces of hydrodynamic o`n; in dry granular systems where only contact forces are present, such structures cannot form for a system in force balance, since they would imply unbalanced forces on the participating particles.   We also note that
our network filtrations (Figure~\ref{fig:2d_filtration}(b)) are formed by the particles and the edges representing the force interaction between them, and hence once a loop forms it never dies
(loops cannot merge).  
The approach that we use here is slightly different from 
the one implemented elsewhere~\cite{physicaD14,pre13,pre14}, where `trivial' loops (formed by three adjacent particles) were not considered.
In the present work, we find that including trivial loops influences the results only marginally, so we keep 
them in our calculations for simplicity.  

On the one hand, persistence diagrams provide a significant data reduction, expressing properties of complex
weighted networks by a collection of points.  However, these diagrams are essentially point clouds, 
and one needs to decide how to extract relevant information.  While 
different approaches have been considered in earlier works~\cite{physicaD14,pre13,pre14,pre16a,pre16b}, 
here we will focus on a single measure, the total persistence, which essentially quantifies how `mountainous' is the interaction
network, with larger distance from peaks (where a structure is born) to a valley (where structures merge and one dies) contributing more to the total persistence measure. 
The total persistence $\tau_\ell$, $\ell = 0$ for components and $\ell = 1$ for loops, 
is defined as the sum of the distances of all points from the diagonal 
of a persistence diagram.  More precisely, 
\[
\tau_{\ell} = \sum_{(b, d) \in PD_{\ell}} {|d - b|}
\]
where we use the value $d = 0$ for the points in the persistence diagram $PD_{\ell}$ that never 
disappear, shown as $(b, -0.1)$ pairs in Fig.~\ref{fig:pers_diags_2d}. 
In superficial terms, one could think of $\tau$ as a measure of how structured the interaction network is; larger total 
persistence means more structure.   Conveniently, choosing $\tau$ as a measure reduces this complex structure
to a pair of scalars.  Clearly, considering total persistence instead of 
persistence diagrams leads to additional data reduction and loss of information.  However, we will show that consideration of total persistence provides insight into the connection of topological properties of the interaction 
networks and rheology of the suspension.   
After discussing total persistence in what follows, we will return to persistence 
diagrams and Betti numbers to discuss how these relate to  properties of interaction networks as the applied stress is varied, in both 2D and 3D. 

\section{Total Persistence as an Order Parameter}
\begin{figure*}
\centering
\subfigure[]{
\includegraphics[width=.45\textwidth]{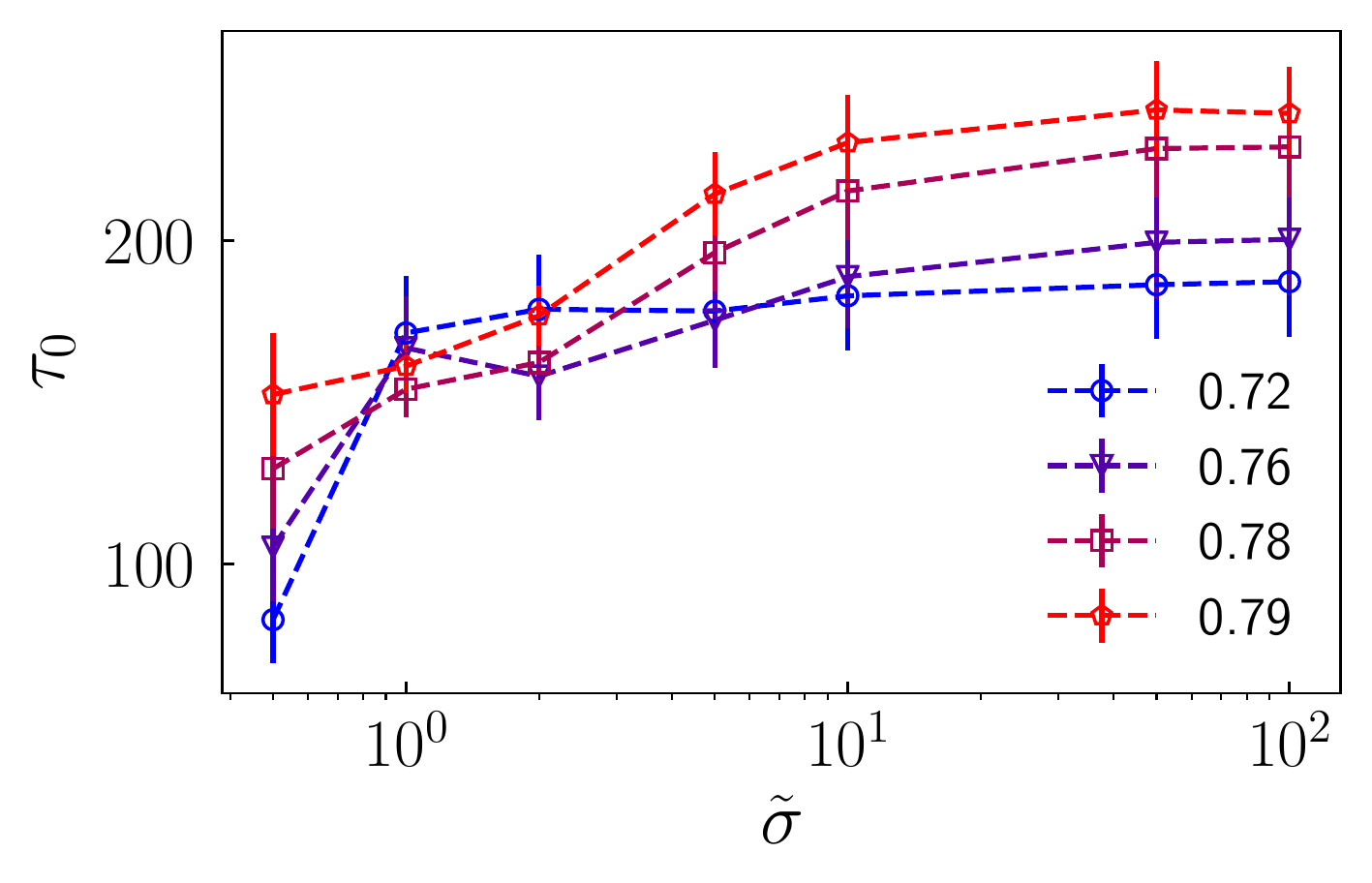}}
\subfigure[]{
\includegraphics[width=.45\textwidth]{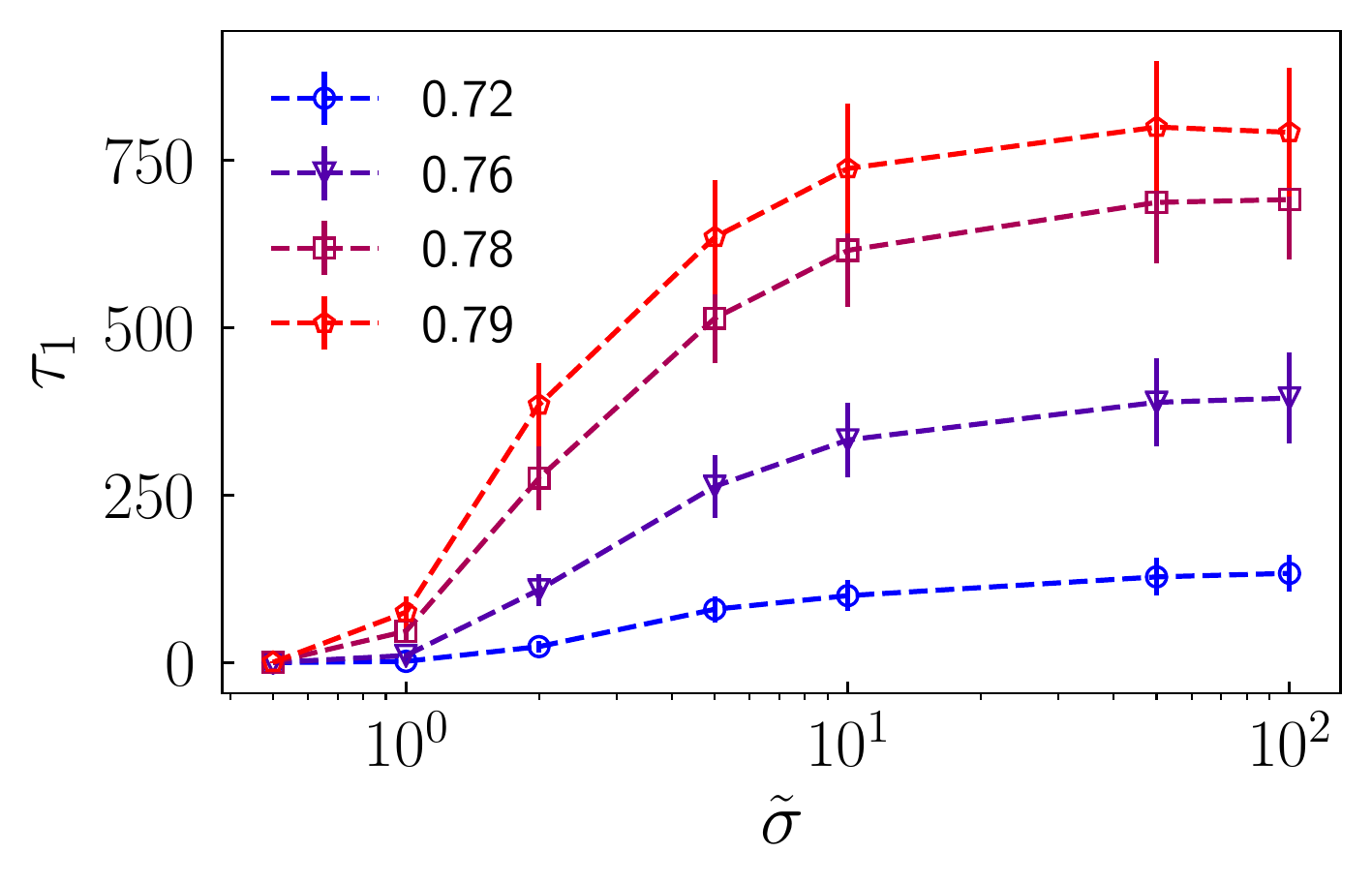}}
\caption{Total persistence for 2D data as a function of $\tilde{\sigma}$ for components (a)
and loops (b), for several values of packing fraction $\phi_{\rm A}$.}
\label{fig:tot_pers_2D}
\end{figure*}

Figure~\ref{fig:tot_pers_2D} shows the total persistence for 2D simulations at four packing fractions as a function of imposed stress. The data are obtained by averaging over 200 
samples of the interaction network at each $(\phi_{\rm A}, \tilde{\sigma})$ condition; we show the mean and the standard deviation.   The total persistence for the connected
structures ($\tau_0$) 
rises strongly in the transition region, from the low viscosity state at $\tilde{\sigma} <1$ to the fully thickened 
one at $\tilde{\sigma}>10$.  The transition is even more distinct for the case of the loop ($\tau_1$) structures, as the 
total persistence is essentially zero for $\tilde{\sigma}<1$ and rises rapidly for $\tilde{\sigma}>1$, again saturating at $\tilde{\sigma} \approx 10$; 
this saturated total persistence for $\beta_1$ is considerably larger for higher $\phi_{\rm A}$, with the value 
for $\phi_{\rm A}= 0.79$ roughly double that at $\phi_{\rm A}  = 0.76$. 
The clear message is that the loop structures are more prevalent 
for the larger $\phi_{\rm A}$, and once formed many of these structures persist for all lower force thresholds: this is as expected on physical
grounds, as the maximum force regions can be separated from one another, but stress continuity in the material 
requires that they be connected by lower force contacts.

The increase of the total persistence $\tau$ with both $\tilde{\sigma}$ and $\phi_{\rm A}$ is directly correlated with the increase in the relative viscosity 
(and as shown elsewhere \cite{mari2014shear}, there is a coincident increase in the normal stresses).  The correlation between
$\tau$ and $\eta_r$ makes the total persistence useful as an  
order parameter that is indicative of a transition in the state of the material. This transition is manifested through the growth of
total persistence as a function of imposed stress for an interaction network of frictional contacts, that form more abruptly and 
become stronger at larger $\phi_{\rm A}$. 

\begin{figure*}
\centering
\includegraphics[width=.4\textwidth]{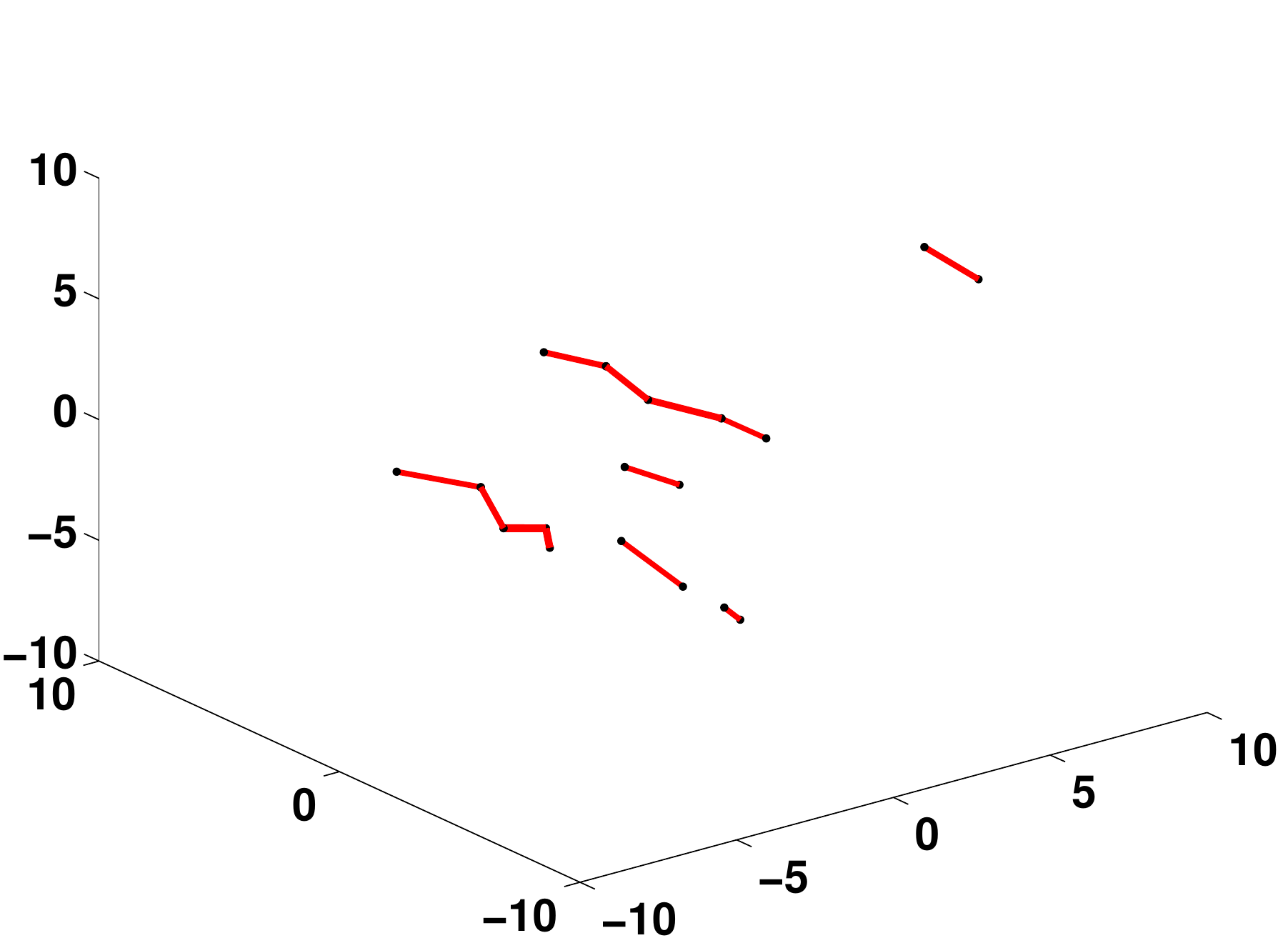}
\includegraphics[width=.4\textwidth]{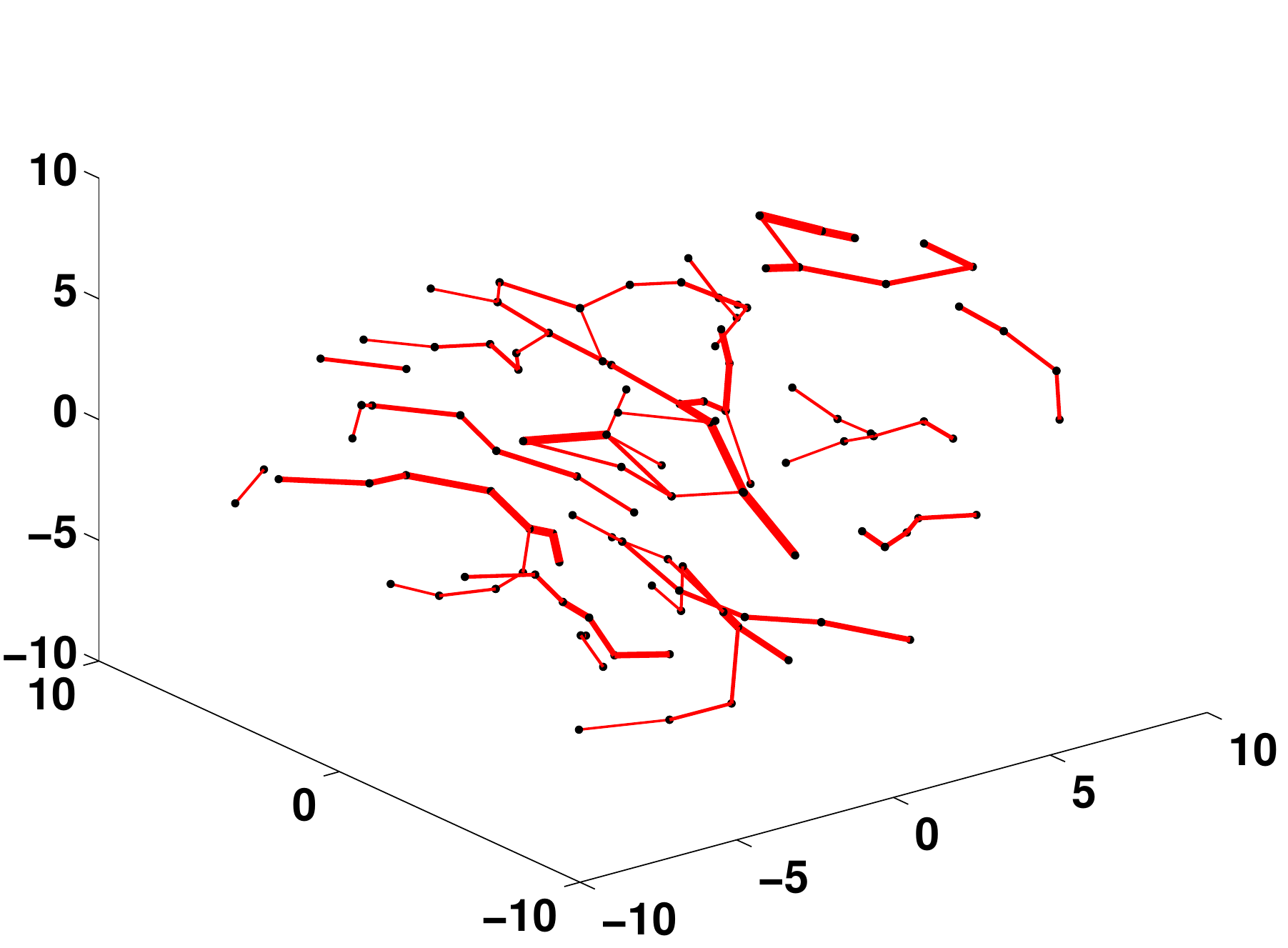}
\includegraphics[width=.4\textwidth]{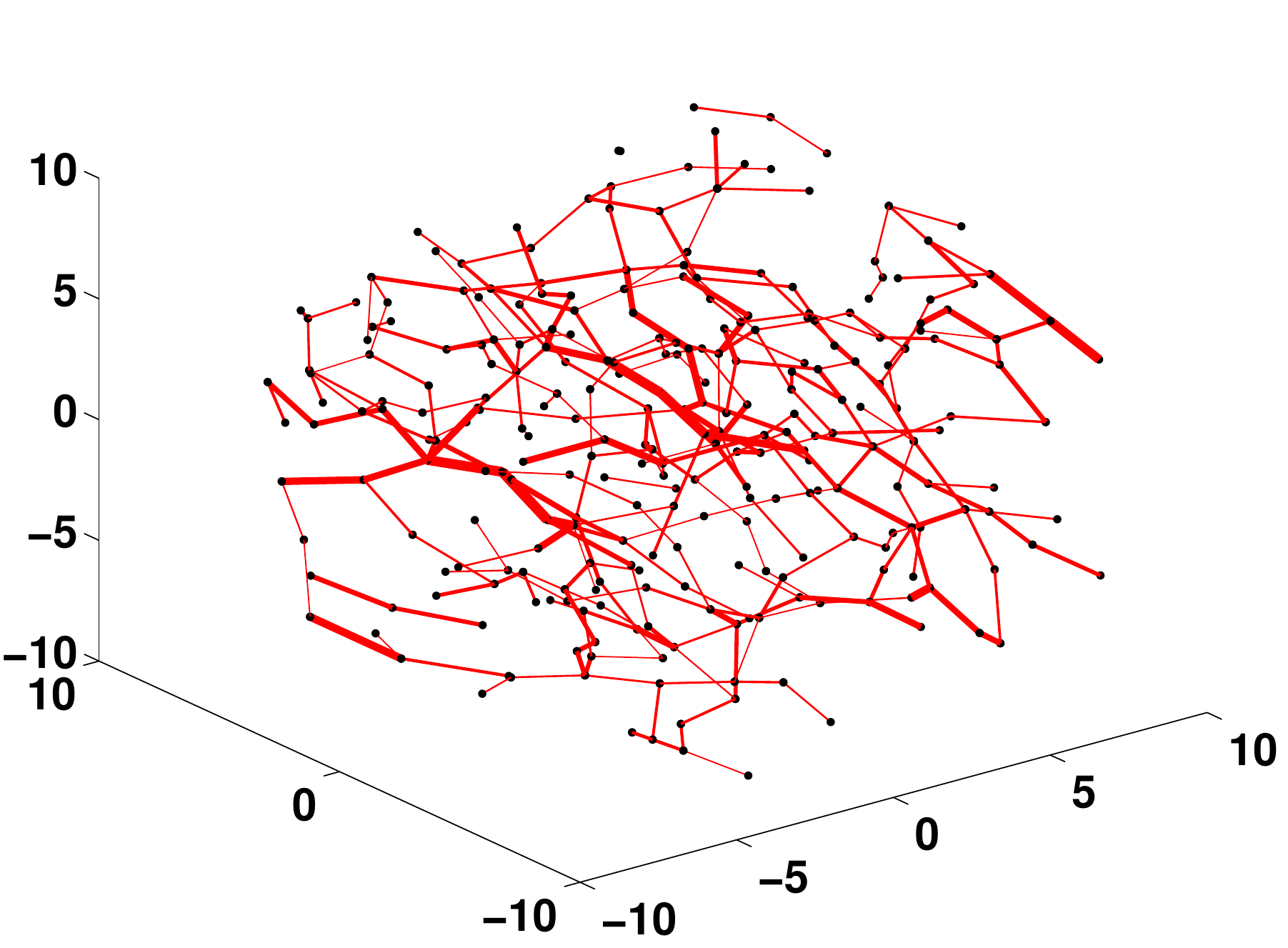}
\includegraphics[width=.4\textwidth]{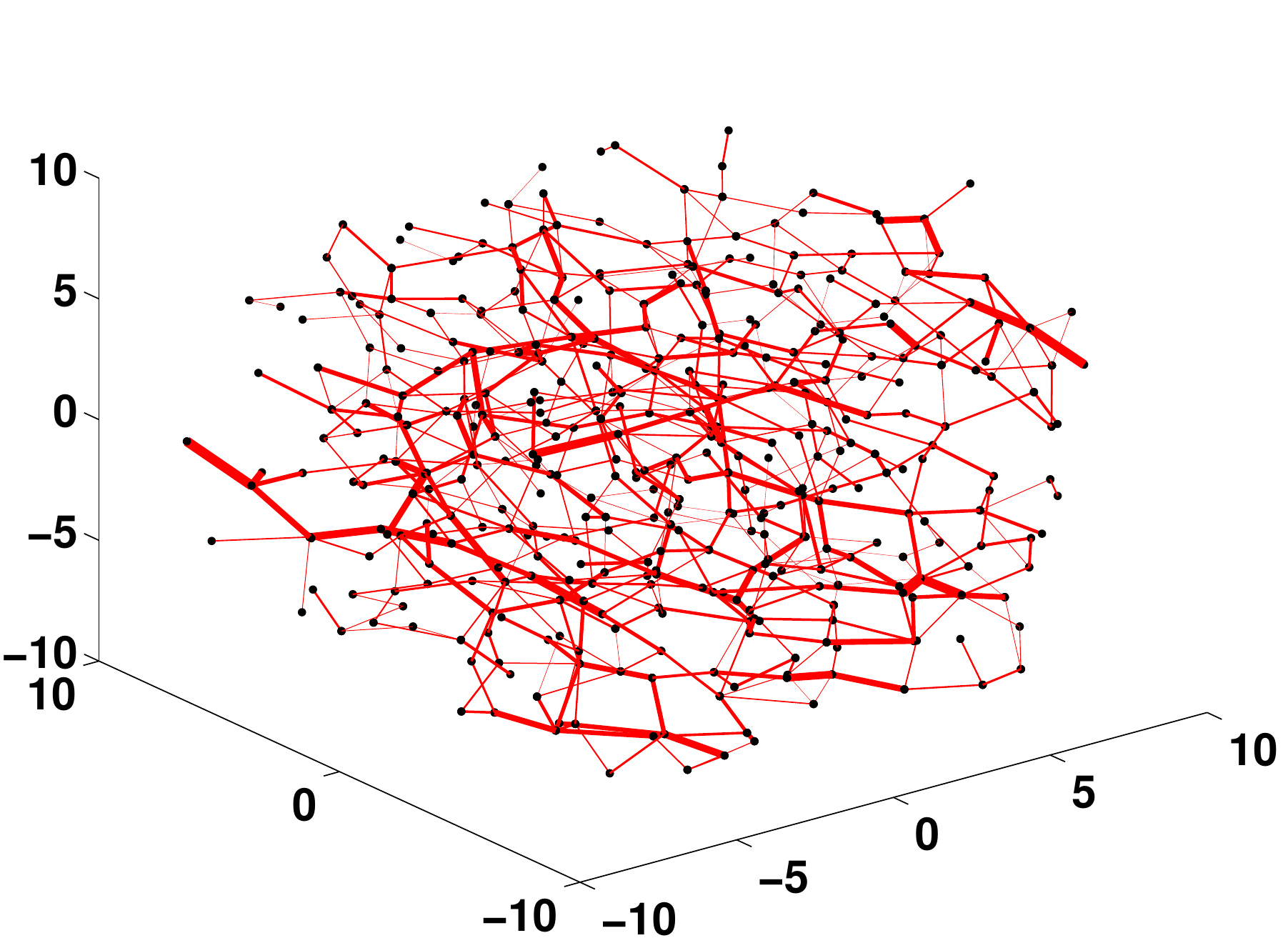}
\caption{Interaction networks for the 3D data for $\phi = 0.56$, for 
$\tilde{\sigma}=0.5$, $\tilde{\sigma}=1$, $\tilde{\sigma}=2$, and $\tilde{\sigma}=10$, each showing a single instant from the simulation once the structure has developed. }
\label{fig:force_chains_3D}
\end{figure*}

This observation is consistent with the picture emerging from the interaction networks displayed as a 
function of increasing stress in Fig.~\ref{fig-3}, or as a function of 
decreasing force threshold in the filtration illustrated by Fig.~\ref{fig:2d_filtration}.   
While the 2D geometry allows ready visual identification
of the network structure, the main interest is ultimately in 
3D simulations.  It is particularly valuable to have quantitative measures in 3D, where visualization is more difficult.  
The simulation approach used in this study has been shown to closely replicate experimental behavior \cite{Mari2015discontinuous}. 
The difficulty of identifying the networks in 3D flows is illustrated in Fig.~\ref{fig:force_chains_3D} by snapshots of the contacting particles (shown as dots) 
and the forces (shown as connecting lines whose thicknesses  indicate the magnitude of the force).   
The configurations are from simulations at $\phi =0.56$ at varying imposed stress from $\tilde{\sigma} = 0.5$ to $\tilde{\sigma} = 10$, taken after sufficient strain, of $O(1)$, has been imposed to allow development of the structure.  
Unlike the equivalent network structures for 2D shown in Fig.~\ref{fig-3}, here we do not show the particles.  Regardless of whether they are shown in a transparent form, the particles 
obscure the view, which even without the particles is clearly less enlightening than the equivalent 2D visualization. 

\begin{figure*}
\centering
\subfigure[]{
\includegraphics[width=.45\textwidth]{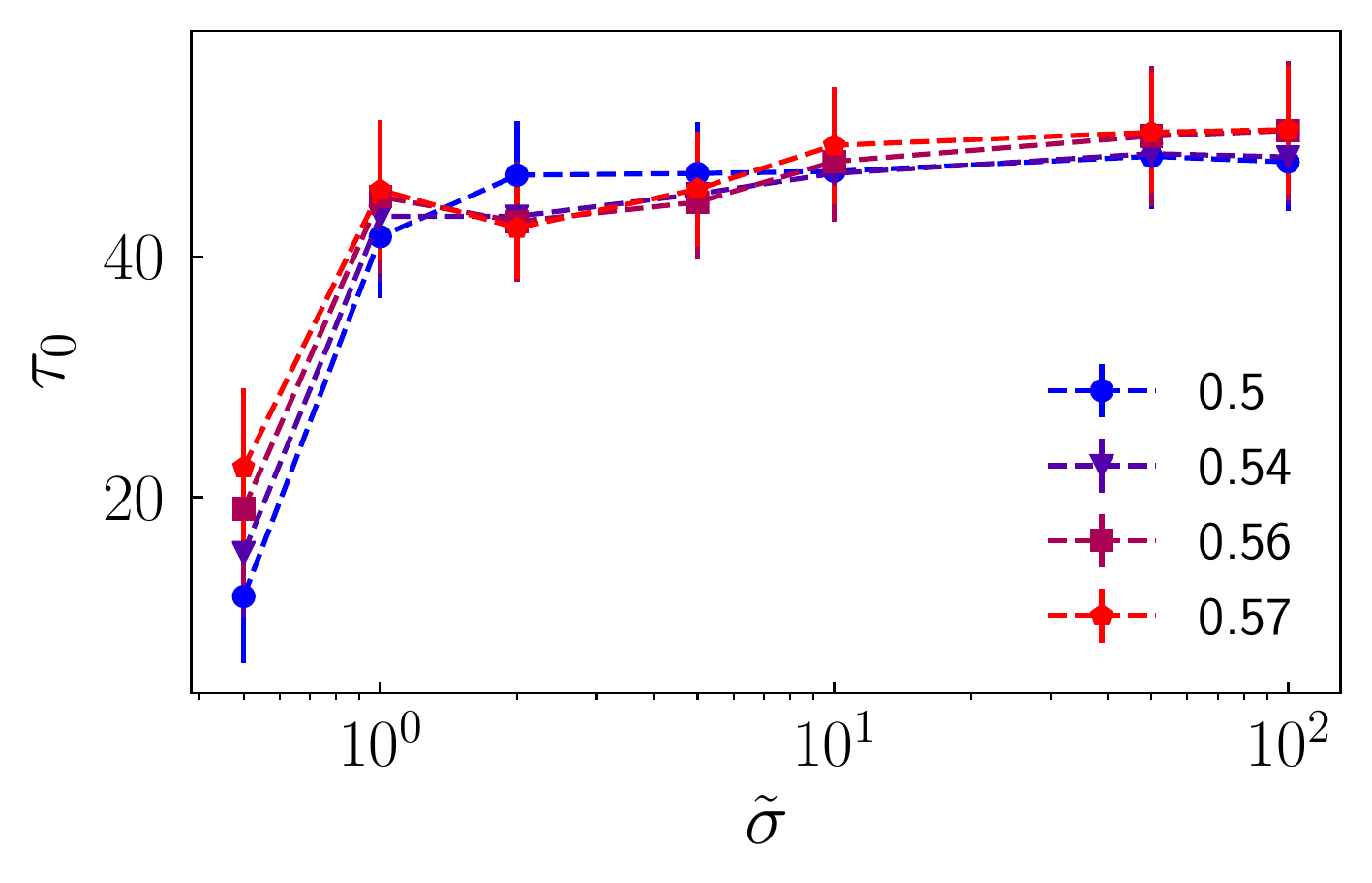}}
\subfigure[]{
\includegraphics[width=.45\textwidth]{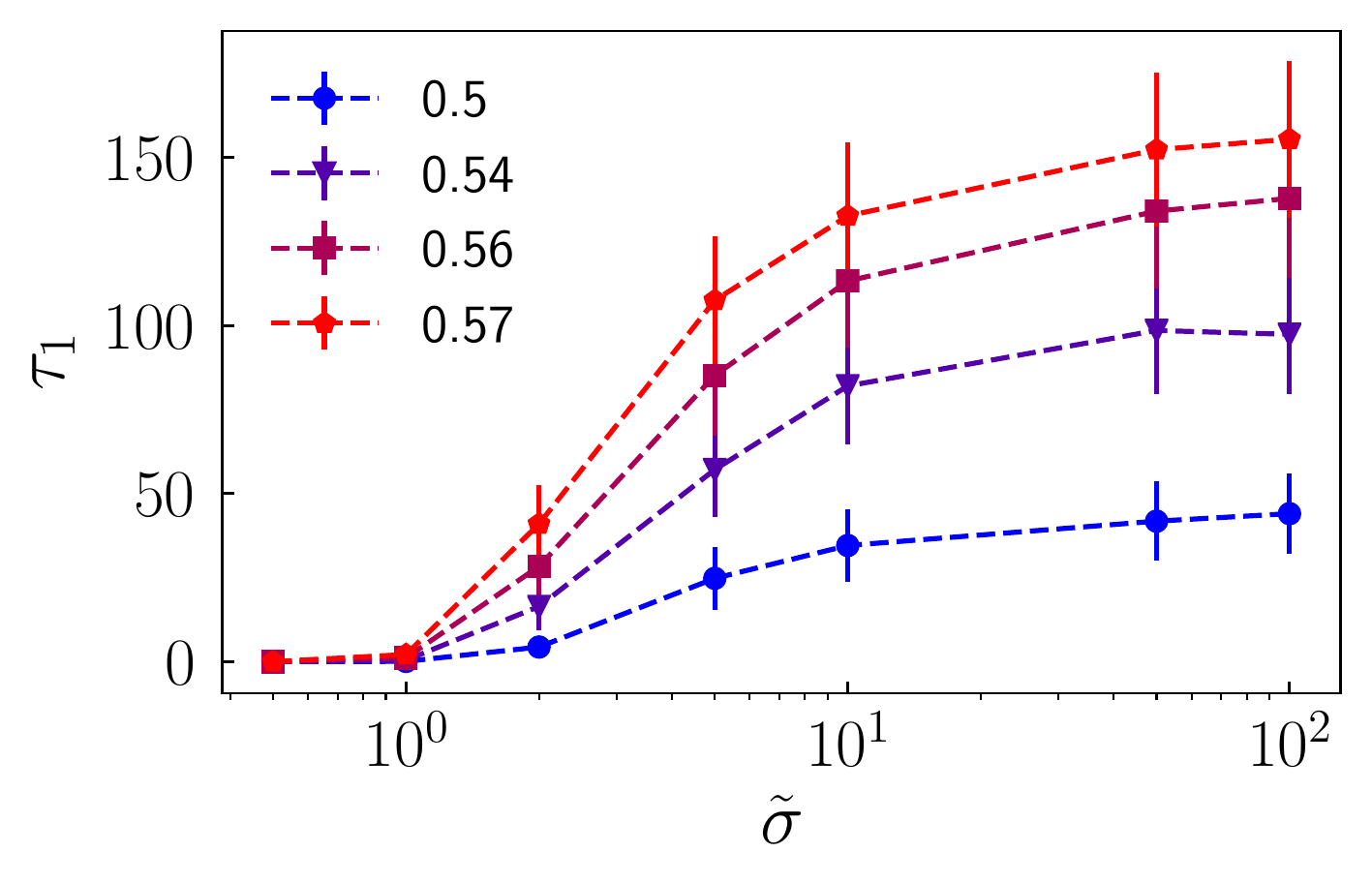}}
\caption{Total persistence for 3D data as a function of $\tilde{\sigma}$ for components (a) and loops (b), for several values of packing fraction $\phi$.}
\label{fig:total_mean_pers_3D}
\end{figure*}

We thus turn to the persistent homology measures obtained from our 3D simulations, considering both the similarities with and differences from 2D. 
Figure~\ref{fig:total_mean_pers_3D} shows the total persistence for several $\phi$.
As stress increases, there is a rapid rise in $\tau_0$ between $\tilde{\sigma} = 0.5$ and $\tilde{\sigma}=1$, with near-saturation at $\tilde{\sigma} = 1$, while the loop structure measured by $\tau_1$ begins to increase at $\tilde{\sigma} \approx 1$ and grows up to $\tilde{\sigma} = 100$, with a slowing of the growth apparent above $\tilde{\sigma} = 10$.
We thus find similarity in the measure of the interaction network structure, represented 
by the total persistence, and the viscosity transition seen in Fig.~\ref{fig-1}.  In 3D, the variation of $\tau$ up to $\tilde{\sigma}=100$ indicates that the interaction network structure continues to develop with increasing $\tilde{\sigma}$ well above the initial transition to the higher viscosity region. 

\begin{figure*}
\centering
\subfigure[]{
\includegraphics[width=.45\textwidth]{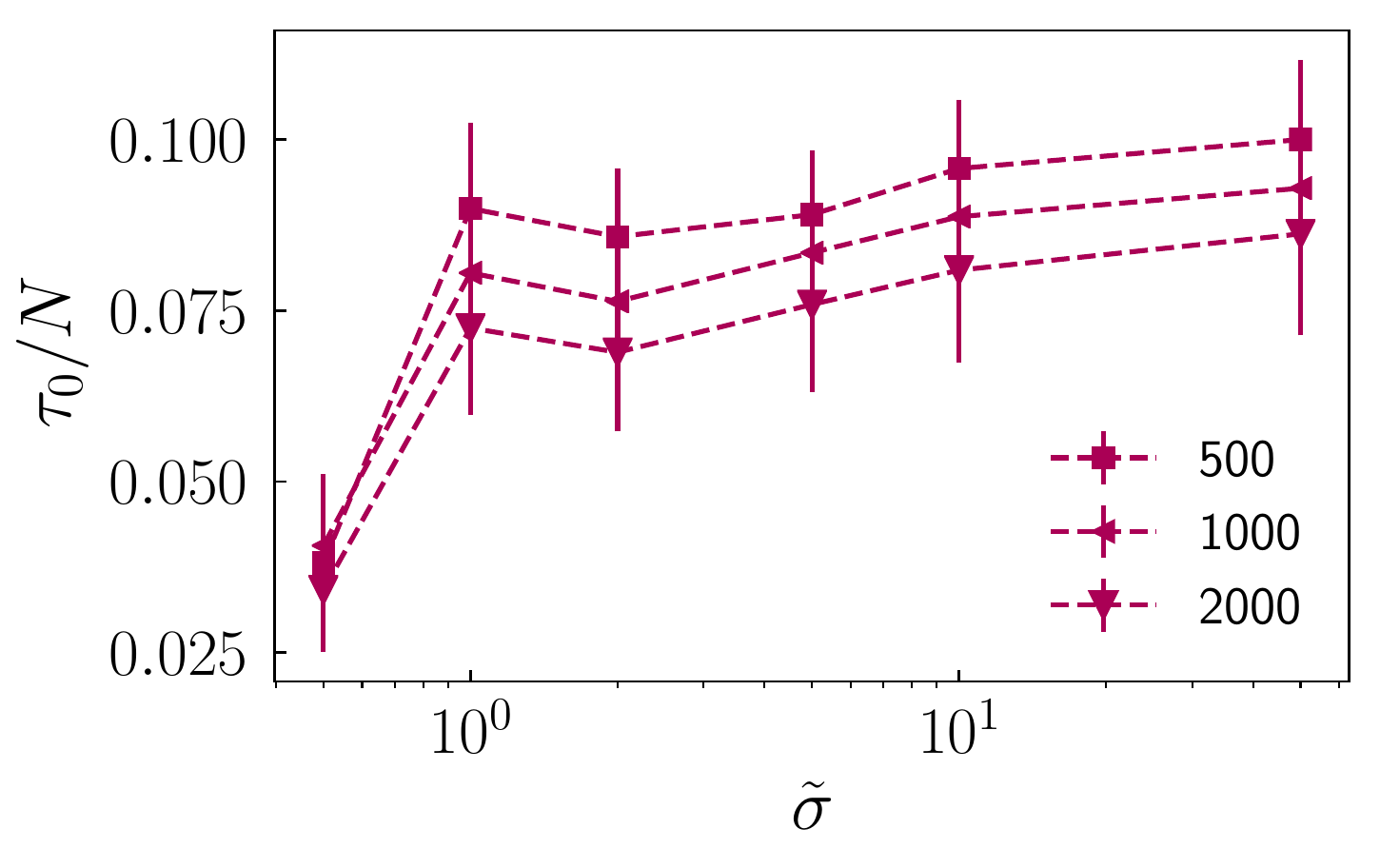}}
\subfigure[]{
\includegraphics[width=.45\textwidth]{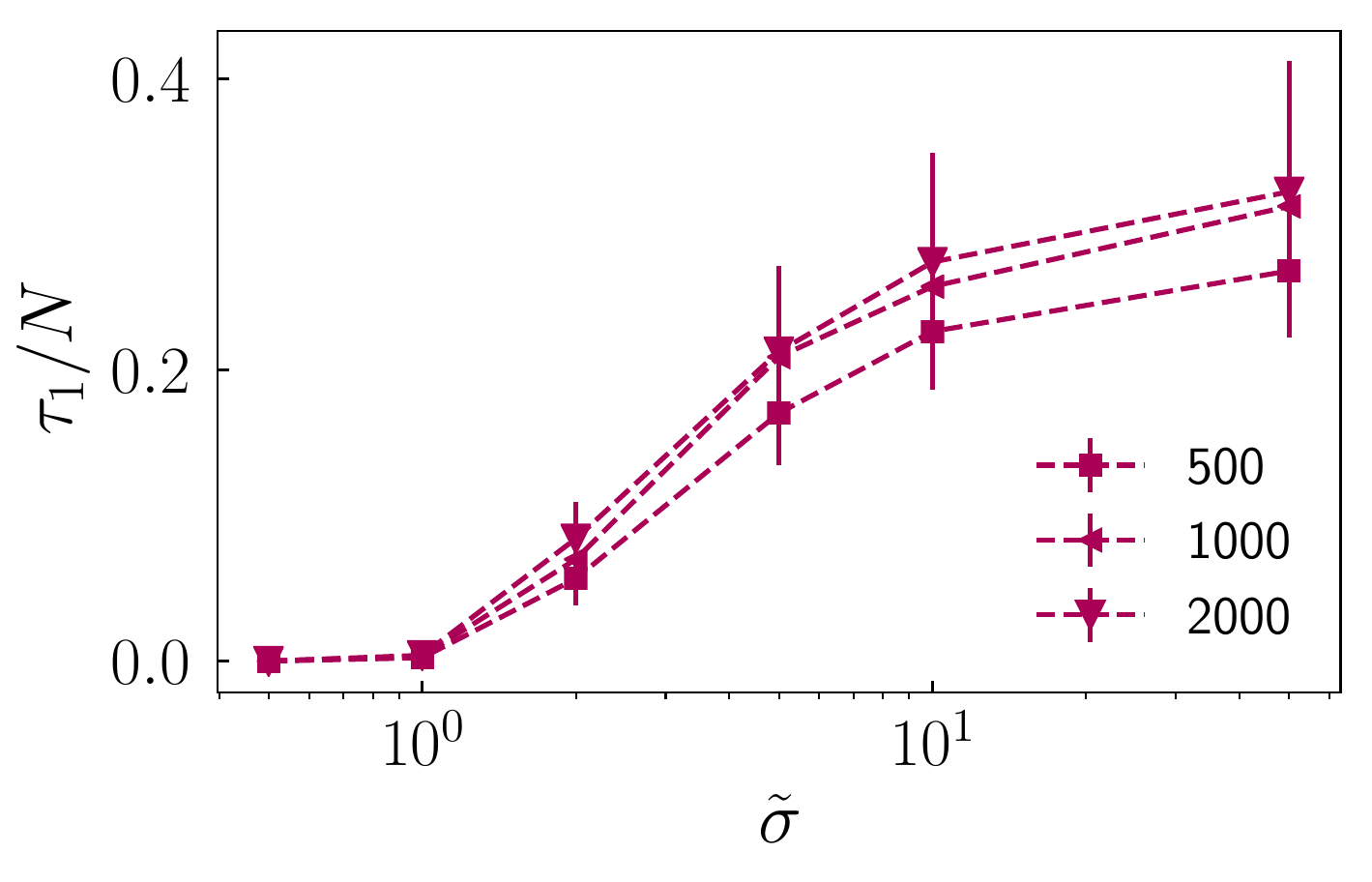}}
\caption{Scaled total persistence for 3D data for $\phi=0.56$ as a function of $\tilde{\sigma}$:
(a) $\tau_0$ and (b) $\tau_1$.}
\label{fig:total_mean_pers_3D_sca}
\end{figure*}

To this point, all results we have presented for 3D were obtained from simulations with $N = 500$.
One obvious question is how the results (for the total persistence in particular) vary with $N$.  To answer this question, we have carried out additional 3D simulations with different numbers of particles. 
Figure~\ref{fig:total_mean_pers_3D_sca} displays $\tau_0$ and $\tau_1$ for 3D simulations at $\phi = 0.56$ scaled by the value of $N$ used for that case.  
We observe the data to collapse within the error bars; however, there does seem to be a weak trend of 
$\tau_0$ decreasing as $N$ increases, and the opposite trend for $\tau_1$.  This trend is explained 
by the boundary effects.  The simulations are periodic in all directions,
but this periodicity is not considered when computing the persistence diagrams. For this reason we miss 
the ``across boundary'' particle interactions. This has the effect of considering some components that are 
supposed to be connected ``across the boundary'' as two separated components, and hence it produces
additional components (points in the diagrams in dimension $0$). Additionally, some loops that are formed 
by connections ``across the boundary'' are not counted as loops, and hence we miss some loops (points in the 
diagrams in dimension $1$). This effect is more noticeable for smaller systems, where the ratio of area
to volume increases.

\begin{figure*}
\centering
\subfigure[]{
\includegraphics[width=.45\textwidth]{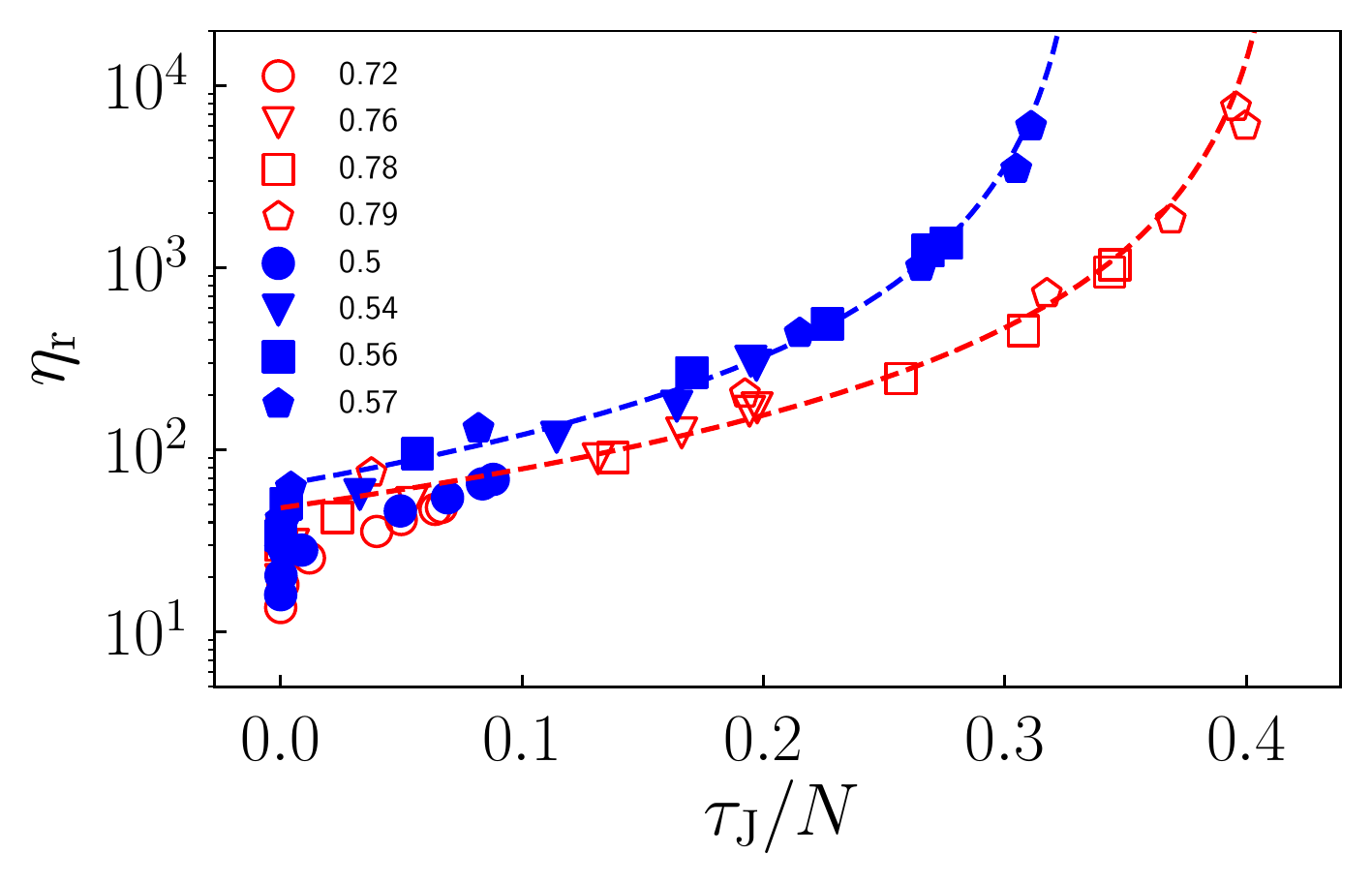}}
\subfigure[]{
\includegraphics[width=.45\textwidth]{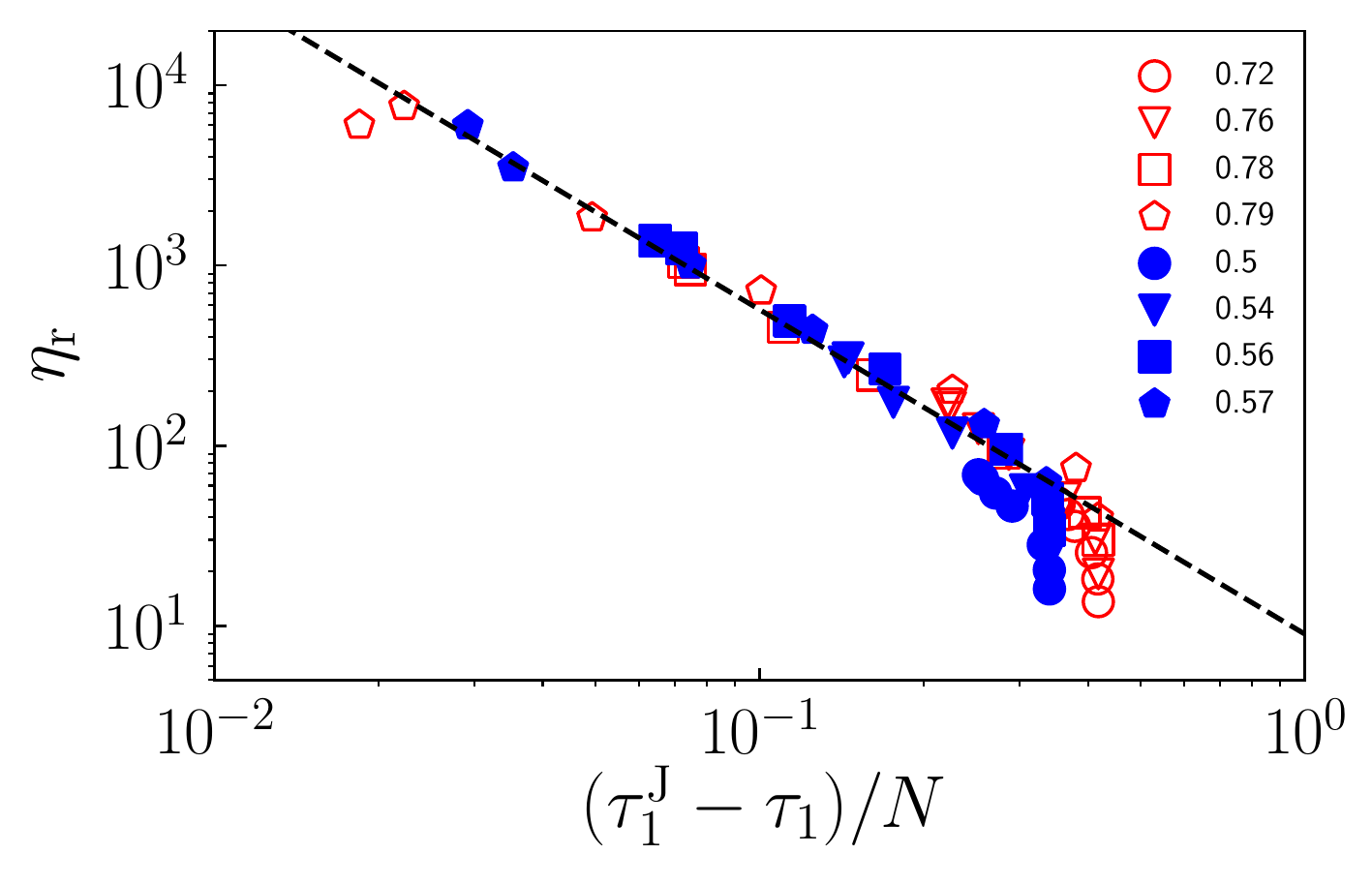}}
\caption{
(a) Relative viscosity $\eta_{\rm r}$ vs. scaled total persistence  (dimension $1$)  $\tau_1/N$ for different packing fractions  for 2 and 3 dimensional data. Solid and empty symbols are used for 2D and 3D, respectively. The dashed lines represent $\eta_{\rm r} \propto ((\tau_1^{\rm J}-\tau_1)/N)^{-\alpha}$ with $\tau_1^{\rm J}/N= 0.34$ and 0.42 for 2D and 3D, respectively, and $\alpha=-1.8$. Red and blue symbols represent two and three
dimensional data.
 (b) The same data on a log-log scale showing $\eta_{\rm r}$ vs $(\tau_1^{\rm J}-\tau_1)/N$.
 The dashed line shows the power law of $-1.8$.
}
\label{fig:total_mean_pers_visc_sca}
\end{figure*}

Figure~\ref{fig:total_mean_pers_visc_sca} displays correlation in both 2D and 3D simulations between relative viscosity and total scaled persistence corresponding to loops, $\tau_1/N$.  This figure shows that the viscosity
 increases with $\tau_1/N$, and appears to diverge at some maximum persistence. The dashed line shows 
$\eta_{\rm r} \propto [(\tau_1^{\rm J}-\tau_1)/N]^{-\alpha}$ where
$\tau_1^{\rm J}/N = \tau_1 (\phi=0.57, \tilde{\sigma}=100)/N =$ 0.34 (for 3D) 
and
$\tau_1 (\phi_{\rm A}=0.79, \tilde{\sigma}=100)/N =$ 0.42 (for 2D) and $\alpha=-1.8$.
The higher--$\phi$ data, in which DST is seen, follow a single scaling law, while for packing fraction $\phi=0.5$ and
$\phi_{\rm A}=0.72$, the data does not agree with the proposed correlation. The structural implication is that the thickened state of DST is associated with a topology of the frictional interaction network  similar to that in the jammed state.

\section{Structure of interaction networks}

Next, we discuss insight to 
structure of the interaction networks that can be gained from consideration of persistence diagrams and Betti numbers. 
We focus in particular on the following questions: (i) Does the structure of interaction networks change 
significantly as shear stress is increased and the system evolves from unthickened to thickened, and does it further evolve as applied stress is increased in the thickened state?   (ii) Is the structure 
different as one progresses from CST to DST at fixed stress by modifying the packing fraction?  (iii) What can be said about the influence of the 
number of physical dimensions on the interaction network structure?

\begin{figure*}
\centering
\subfigure[]{
\includegraphics[width=0.35\textwidth]{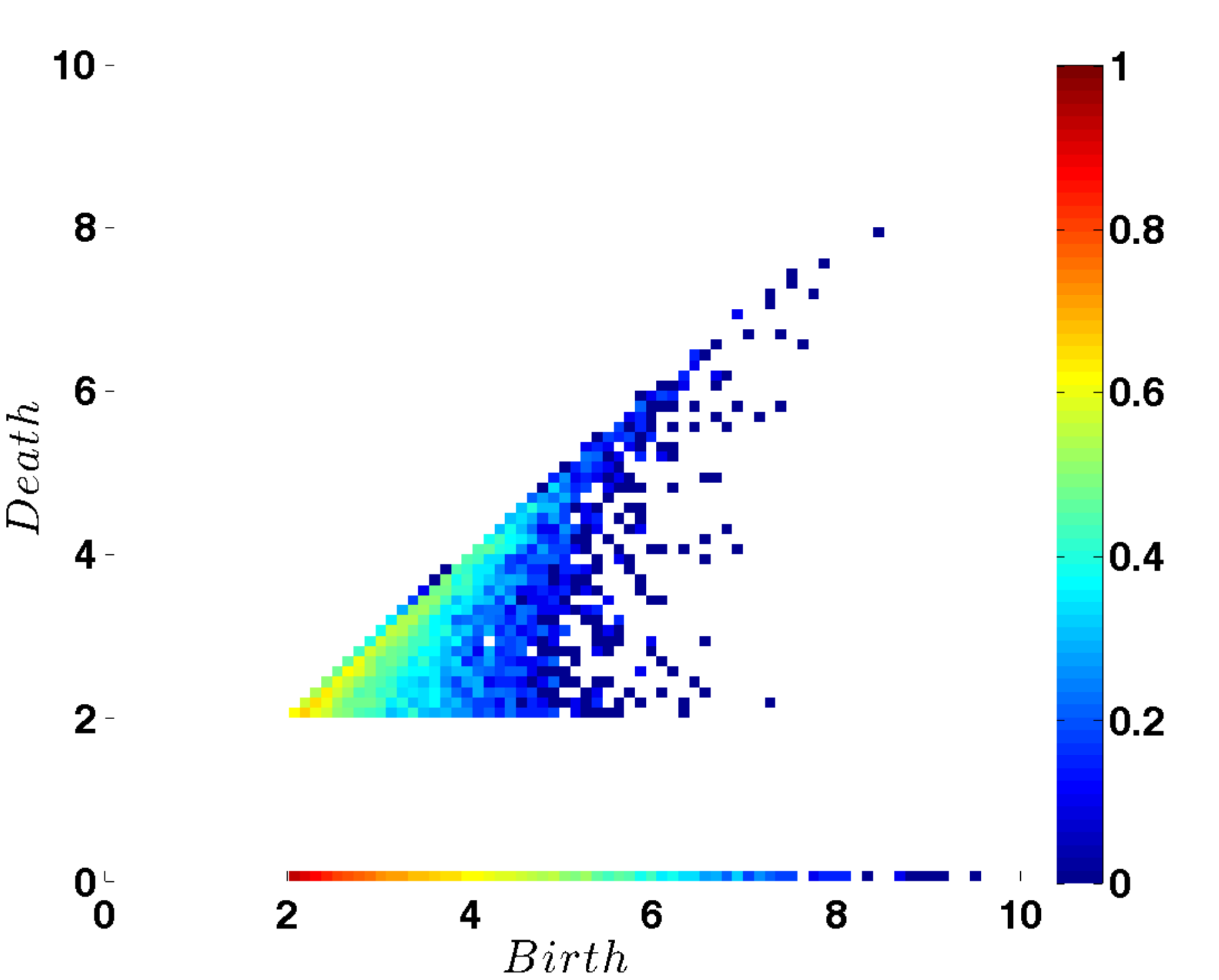}}
\subfigure[]{
\includegraphics[width=0.35\textwidth]{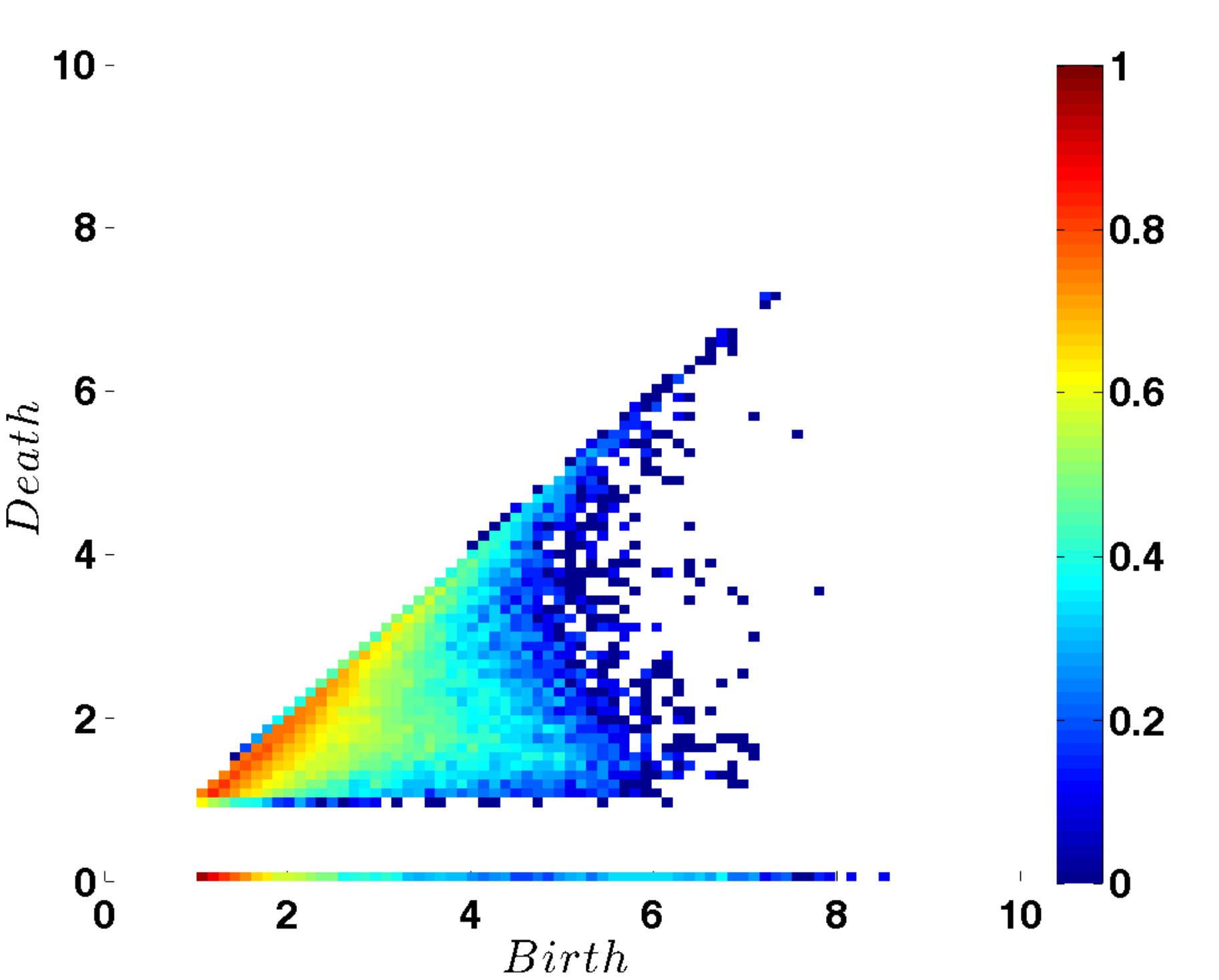}}
\subfigure[]{
\includegraphics[width=0.35\textwidth]{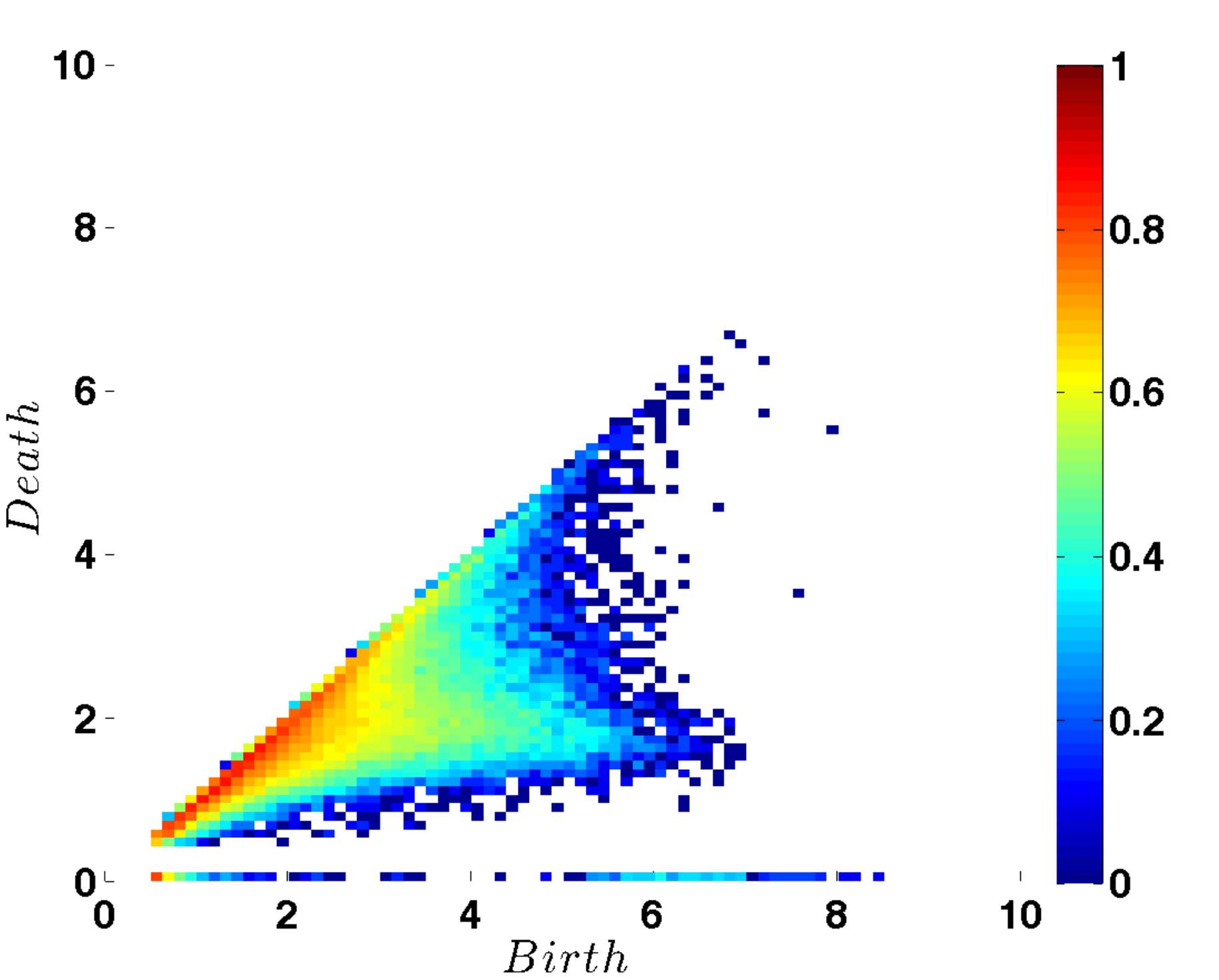}}
\subfigure[]{
\includegraphics[width=0.35\textwidth]{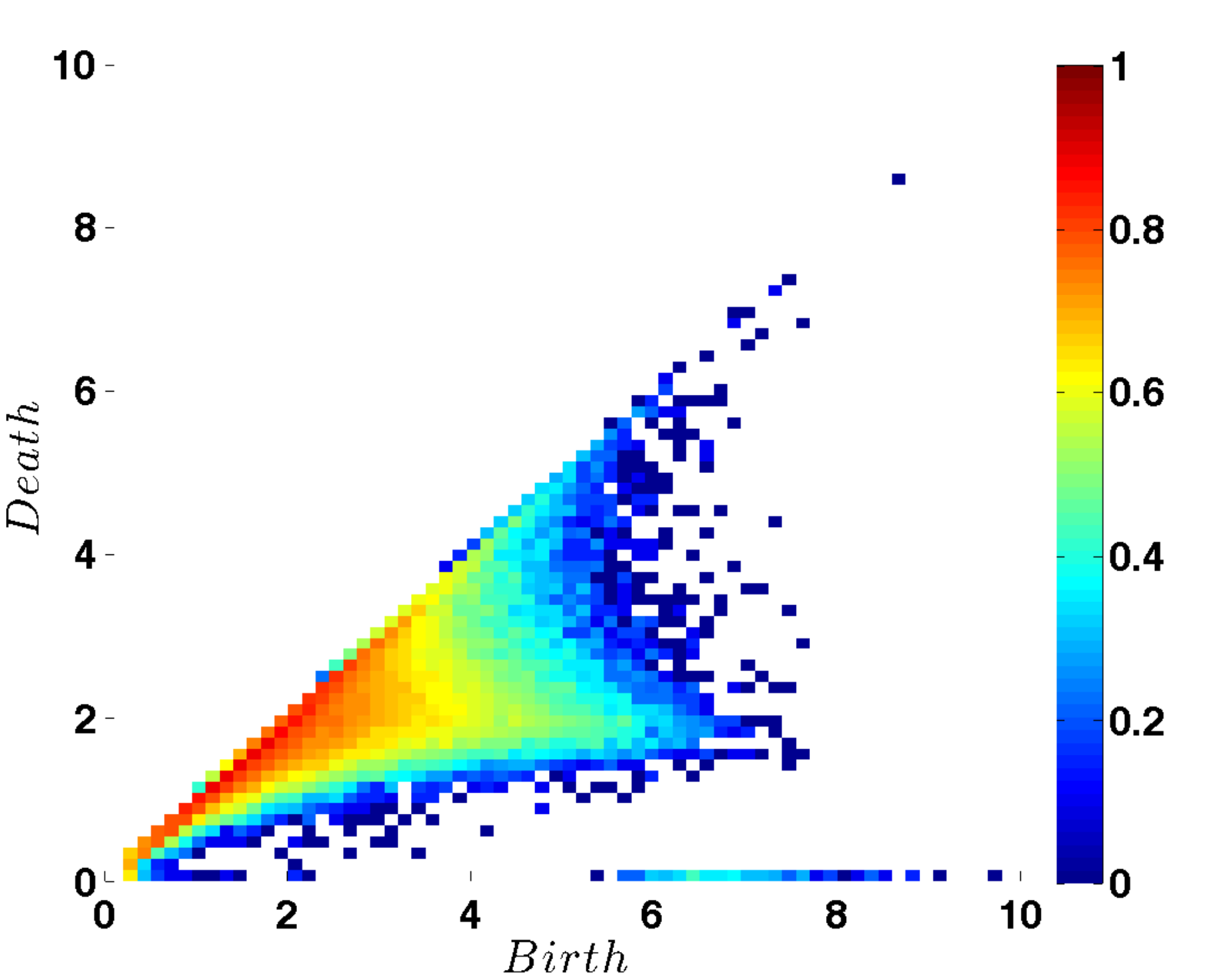}}
\subfigure[]{
\includegraphics[width=0.35\textwidth]{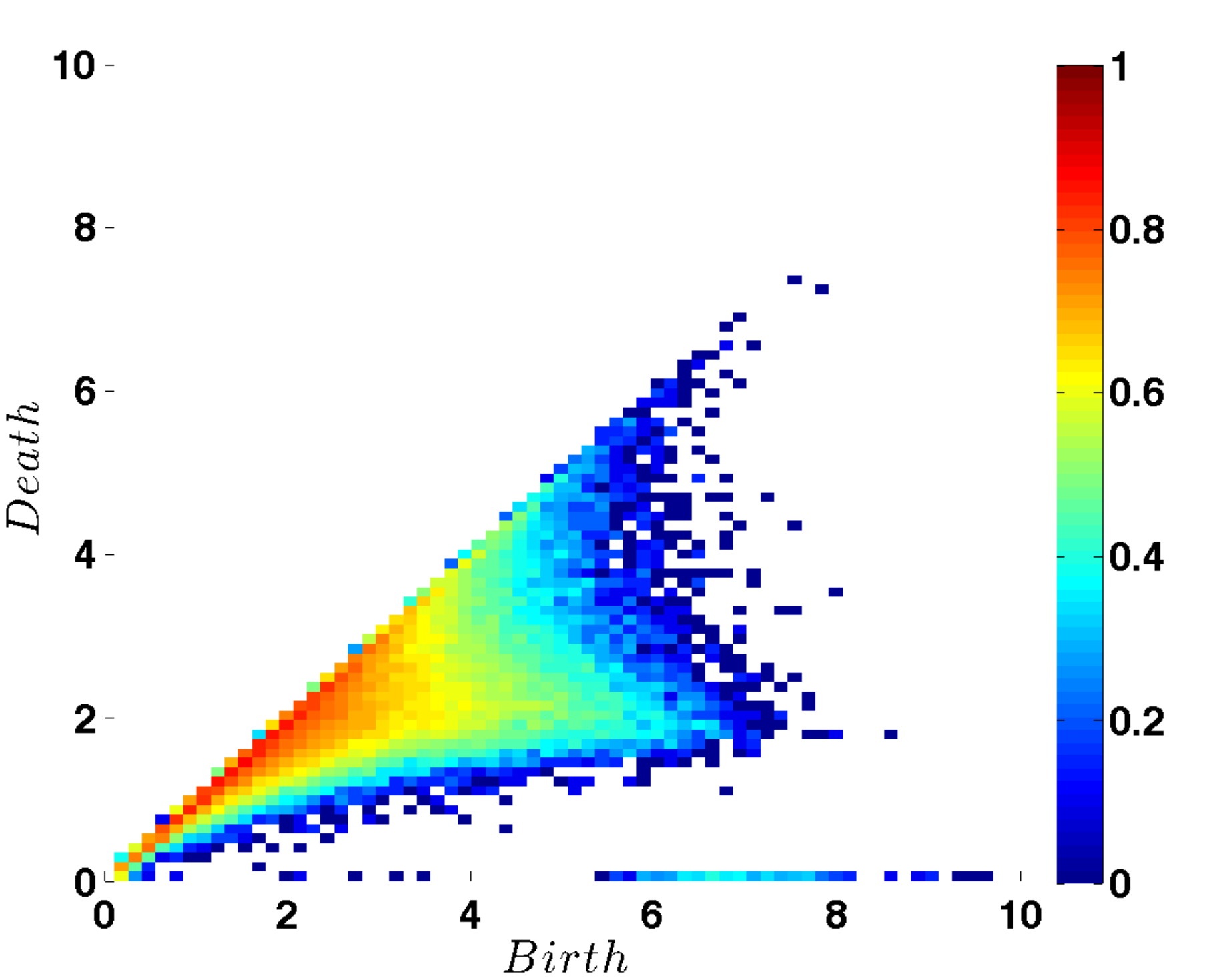}}
\subfigure[]{
\includegraphics[width=0.35\textwidth]{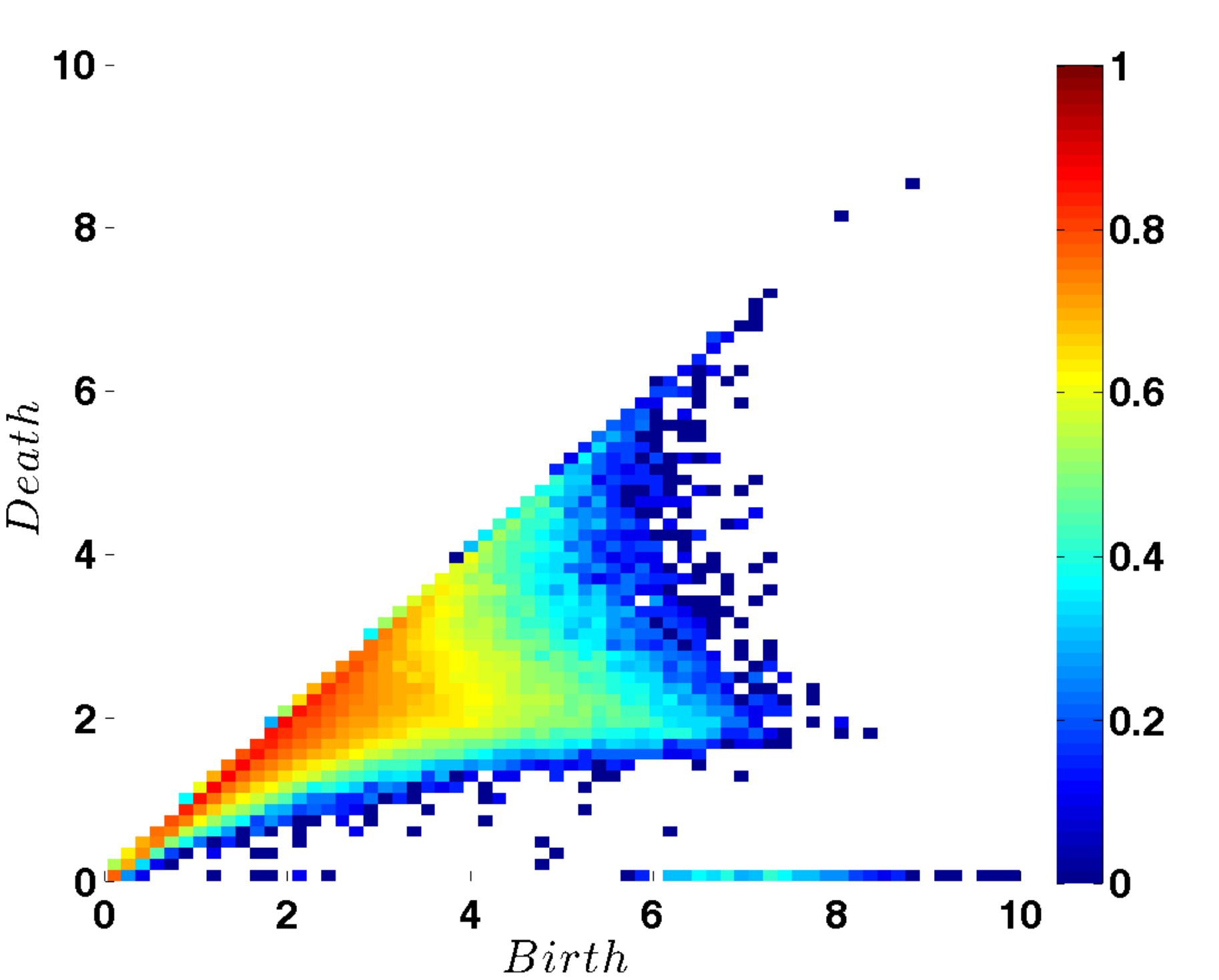}}
\subfigure[]{
\includegraphics[width=0.35\textwidth]{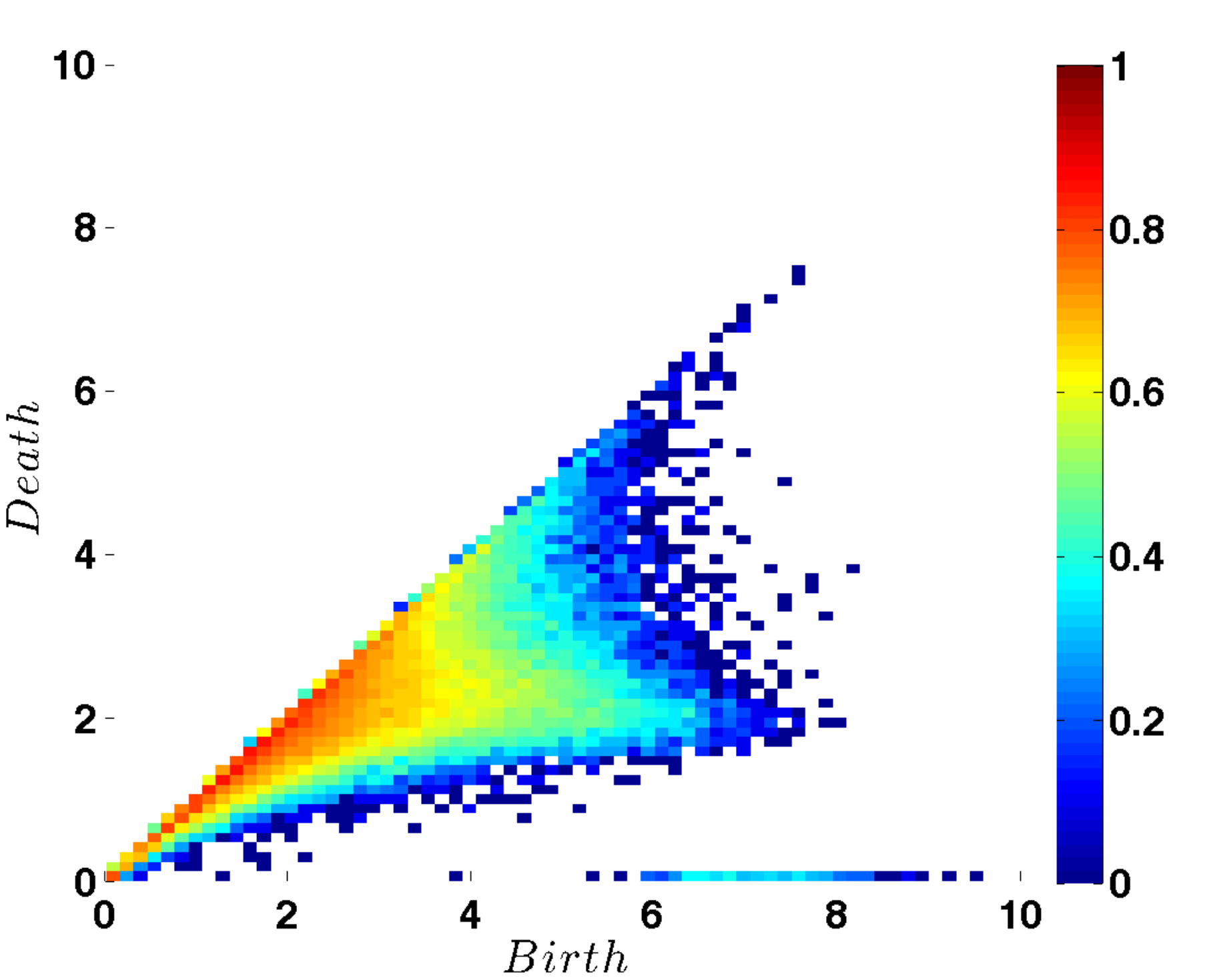}}
\caption{Density plots of all the $\beta_0$ persistence diagrams superimposed for the given parameters values for the for 2D simulations colored according to the density of points. The
color bar indicates the density normalized across all the plots. The plots correspond to $\phi_{\rm A}=0.79$ and for (a): $\tilde{\sigma}=0.5$, (b): $\tilde{\sigma}=1$, (c): $\tilde{\sigma}=2$, (d): $\tilde{\sigma}=5$, (e): $\tilde{\sigma}=10$, (f): $\tilde{\sigma}=50$, (g): $\tilde{\sigma}=100$.}
\label{fig:density_2D_B0}
\end{figure*}

To answer (i) we make use of the persistence diagrams.
As total persistence, $\tau$, compresses all available information into a pair of numbers, the additional information in the persistence diagrams from which these numbers are determined may provide insight.
Figure~\ref{fig:density_2D_B0} shows the superposition of all $\beta_0$  persistence diagrams at each of seven values of imposed stress for the 2D simulations at $\phi_{\rm A}=0.79$.
We observe that as $\tilde{\sigma}$ increases, the density of the points increases.
This is consistent with Fig.~\ref{fig:tot_pers_2D}(a), and we see that once the thickened state is reached at $\tilde{\sigma} \approx 5$, the density of the points shows little further variation. 
Thus, we deduce that as the suspension reaches the thickened state, its primary structural properties are essentially unchanging, and thus the interaction forces on the contacts simply increase proportionally with the applied stress.    This conclusion is supported by considering $\beta_1$ persistence diagrams (figure not
shown for brevity).  

To answer question (ii), observe that if there are significant changes in the structure of the interaction networks, then we expect to see differences in the corresponding persistence diagrams.
Examination of plots such as those shown in Fig.~\ref{fig:density_2D_B0} but generated by changing packing fraction
show only minor differences, consistently with 
total persistence results. 
Essentially, the only influence of packing fraction is a slight shift of the points in persistence diagrams to lower values for smaller packing fractions. 
Therefore, if the system is already in the thickened state, we do not find qualitative differences in the structure of interaction networks for the systems that are brought to thickened state by increasing either applied stress or packing fraction. 

\begin{figure*}
\centering
\subfigure[]{
\includegraphics[width=0.4\textwidth]{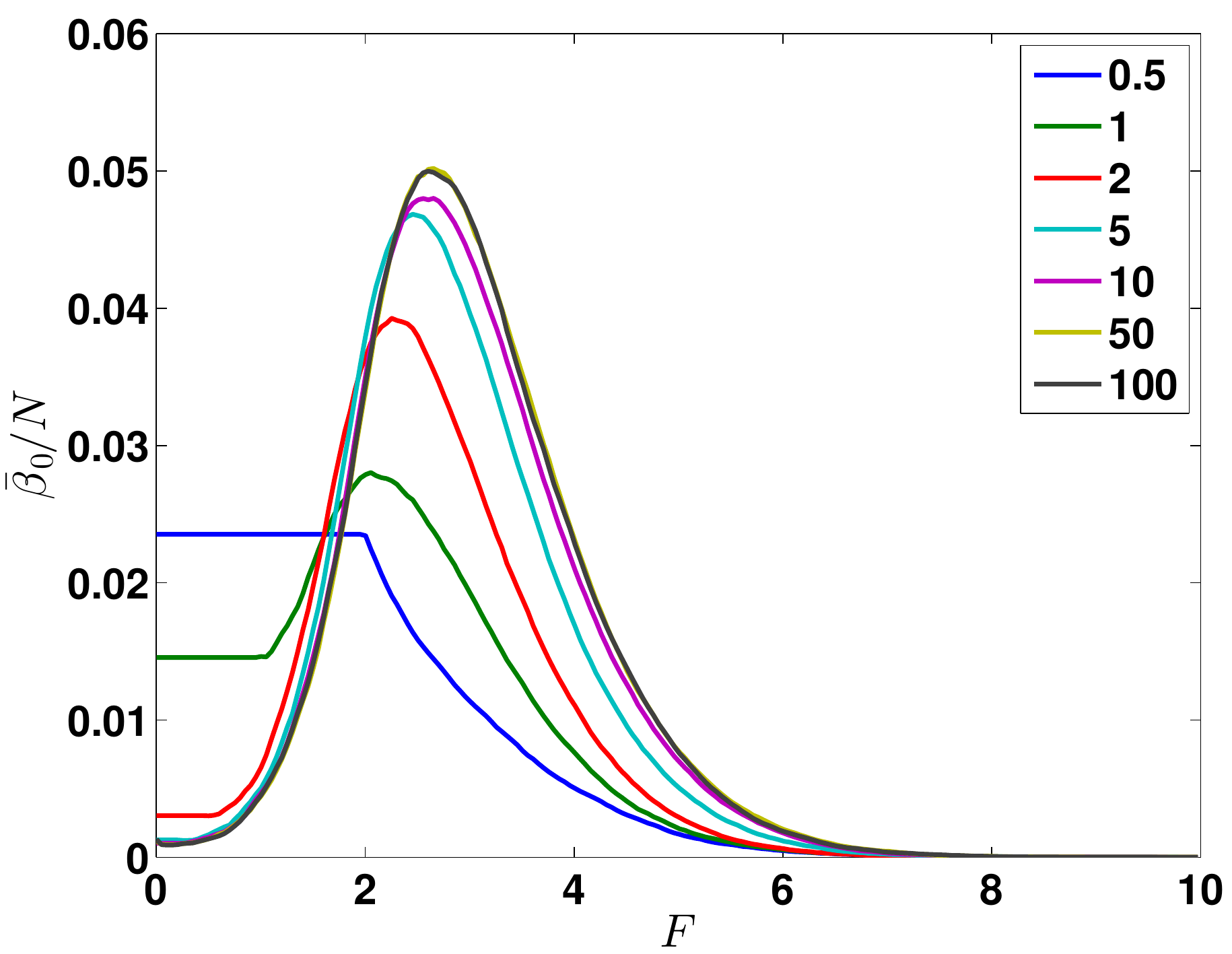}}
\subfigure[]{
\includegraphics[width=0.4\textwidth]{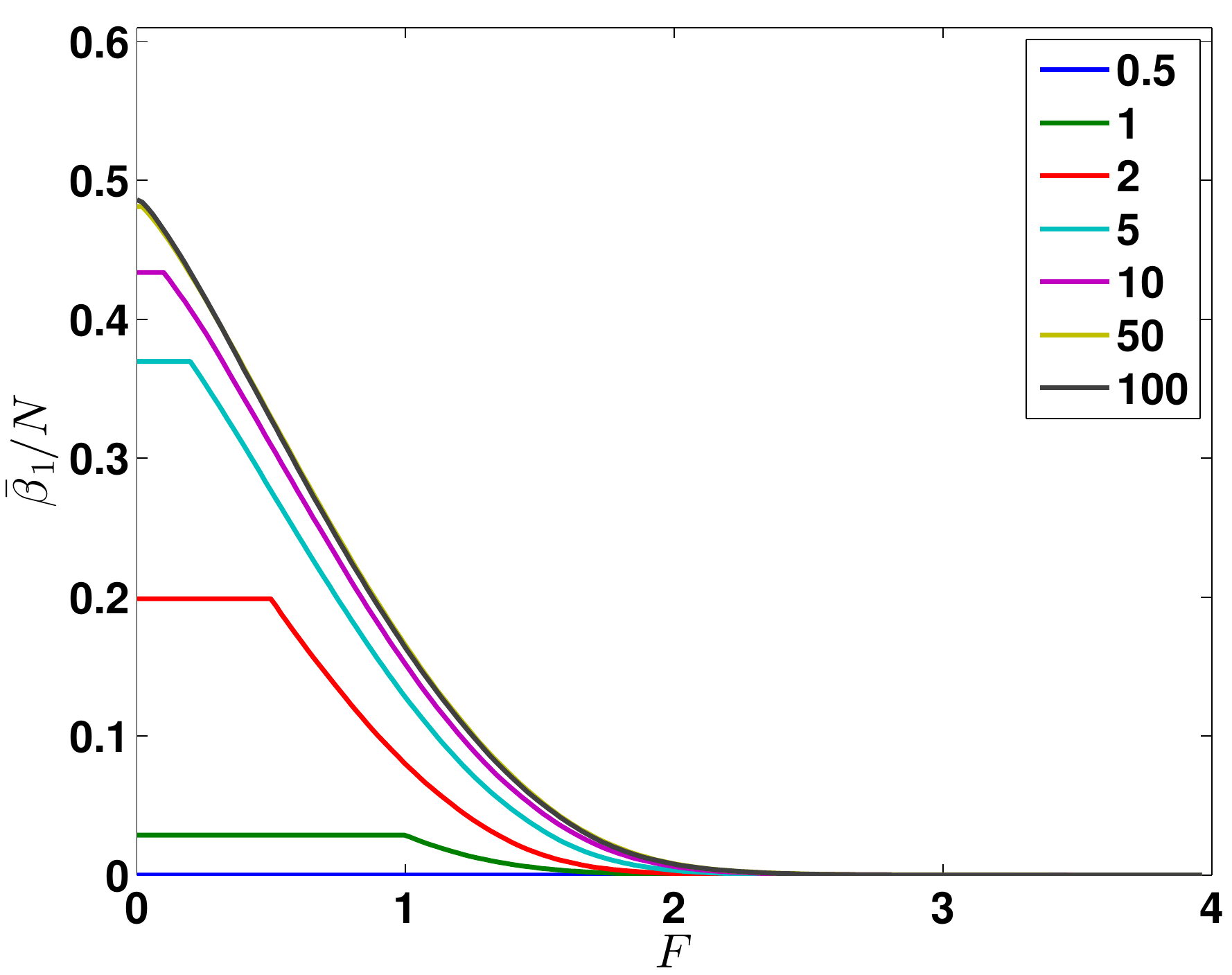}}
\subfigure[]{
\includegraphics[width=0.4\textwidth]{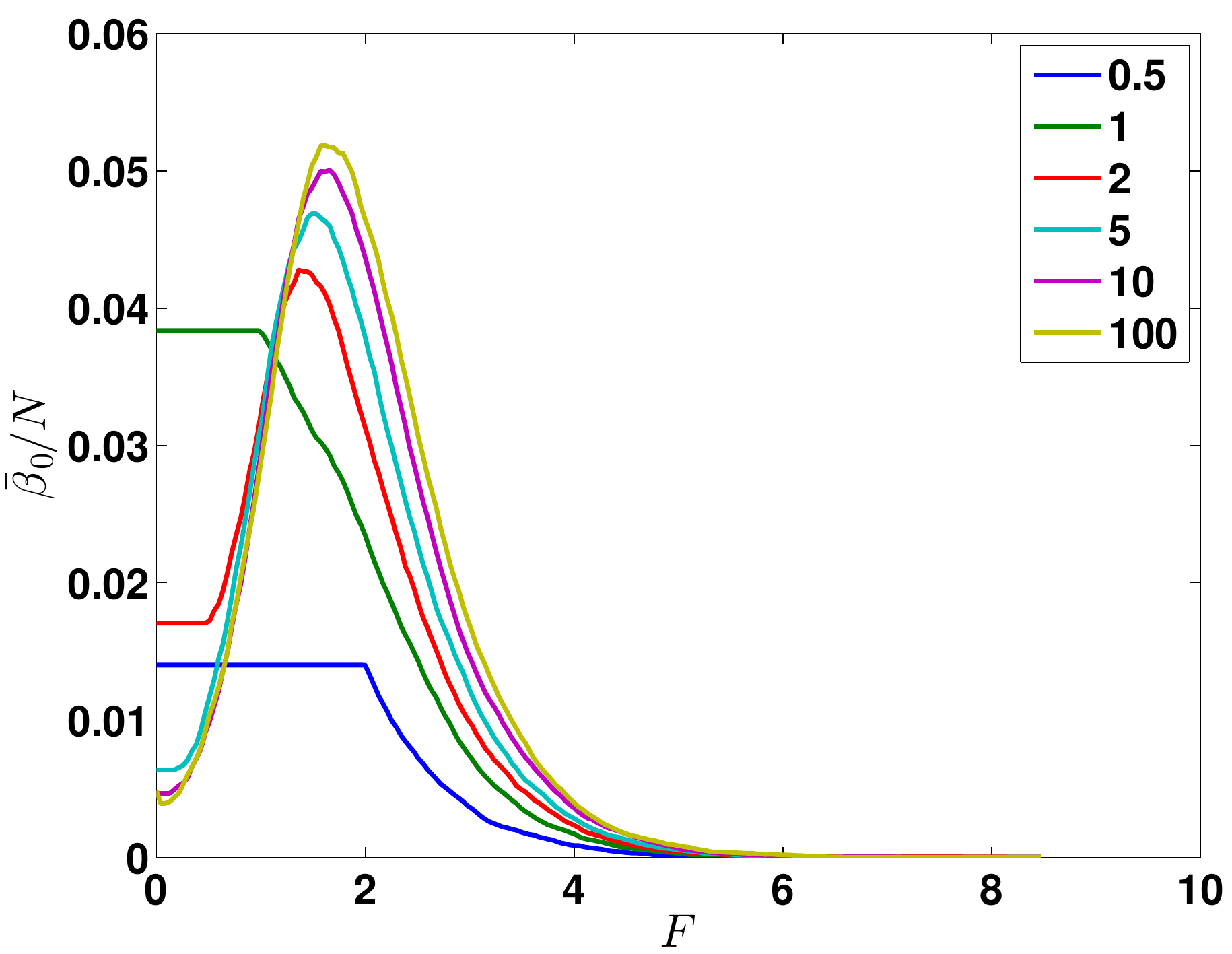}}
\subfigure[]{
\includegraphics[width=0.4\textwidth]{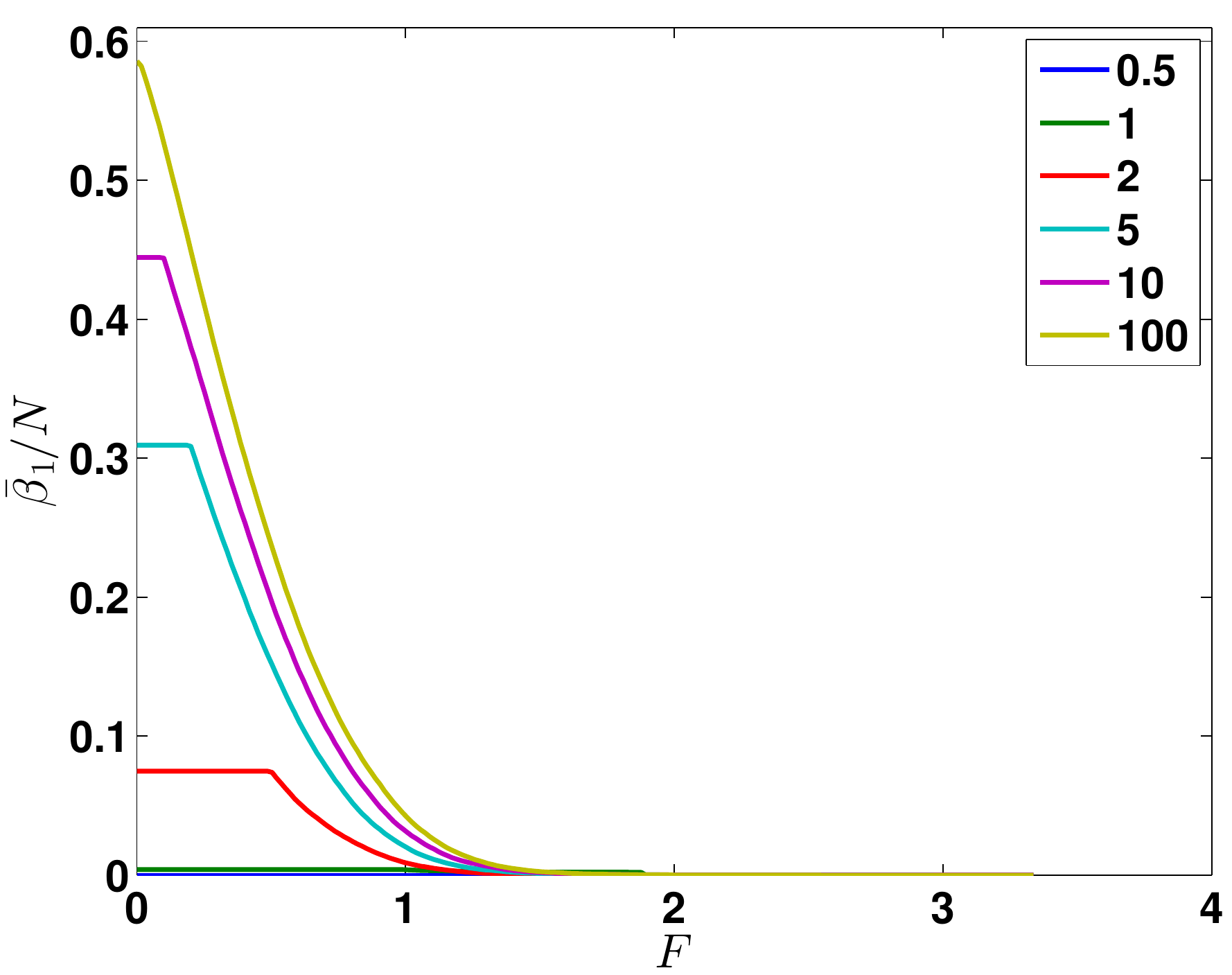}}
\caption{Betti numbers for several values of $\tilde{\sigma}$, scaled by the number of particles $N$ for 2D simulations (a-b) for $\phi_{\rm A}=0.79$, 
and for 3D simulations (c-d) for $\phi=0.56$.  Parts (a, c) show $\beta_0$, and (b, d) show $\beta_1$.
}
\label{fig:betti_2D_3D}
\end{figure*}

Next we address the final question regarding  whether and how the structure of interaction networks depends on the number of physical dimensions.
We note that answering such a question by considering purely topological measures is difficult, since the significant data reduction involved in computing homology may also remove relevant information about the structure. 
Nevertheless, to gain some basic insight, we default first to Betti numbers.  Figure~\ref{fig:betti_2D_3D} shows the Betti numbers for the 2D and 3D simulations, 
scaled by the number of particles.   We find that the Betti number curves are similar for 2D and 3D, especially for 
$\beta_0$ which is strikingly similar for the two dimensionalities at large stress, where the network structure 
has saturated; note that the value of peak shifts to larger $F$ for 2D.  The finding that Betti numbers are so
similar for 2D and 3D is surprising, given the difference in number of near neighbors in 2D (about six) and 3D 
(about 12). 
If we now consider the relationship of total persistence to relative viscosity shown in Fig. \ref{fig:total_mean_pers_visc_sca}, we find a similar scaling law for either 2D or 3D on approach to jamming.  This is again surprising and intriguing, as the total persistence provides a compact description of the connectivity of the network, and this would a priori be expected to differ depending on dimensionality. Such expectations arise from noting that a mechanically stable static or jammed packing depends systematically on dimensionality $D$, varying from 
$Z=2D$ to $D+1$ as the friction coefficient varies from $\mu =0$ to $\mu \rightarrow \infty$ \cite{Liu_2010}.  Thus, we have identified similarities across dimensionality in both the 
Betti numbers for the high-stress, or `fully-networked' conditions, and for the near-jamming total persistence of sheared suspension contact networks.  This similarity between measures of 2D and 3D networks points to a need for a more thorough understanding of the relation of the network structure to the mechanical properties, as it may reveal principles of 
force organization in amorphous materials.  

As a final note regarding Fig. \ref{fig:betti_2D_3D}, the flat portion of the $\beta_0$ and $\beta_1$ curves 
at smaller values of $F$ and small  $\tilde{\sigma}$ are due to hydrodynamic forces that allow for existence
of a non-zero number of components. For large values of $\tilde{\sigma}$ these separate components are not present
since the applied stress is strong enough to make all components connect.  A similar argument explains constant values of
$\beta_1$ for the same values of force threshold. 

\section{Conclusion and perspective}

In this paper we have shown that there is a clear connection between the local properties of the interaction networks 
that develop in sheared suspensions and their global rheological response.  This finding suggests that 
proper understanding of the properties of the interaction networks may be the key in quantifying
their rheological response not only for relatively simple flow geometry considered here, but in other more 
complex settings as well.   
It is encouraging that a relatively simple measure, the total persistence describing structure of loops in the 
interaction networks, appears as an order parameter that describes well the rheology.  This finding is important, 
since it suggests that the details of the interaction networks may be absorbed  into a compact measure for the purpose of understanding 
the rheology.  On the other hand, it should be noted that the total persistence includes (in compressed form) the 
information about interaction network properties for all interaction strengths at once, and therefore understanding and 
quantifying particle-particle interactions may be needed for the purpose of describing the material response. 

One particularly interesting finding is that the total persistence {\it of loops in interaction networks} is the 
measure that correlates the best with the rheological properties (viscosity).  This result suggests
that future research should focus in more detail on the connectivity of the interaction networks in order to understand
better the rheology.  It remains to be seen whether the fact that the total persistence of loops in particular appears
as an order parameter describing the system behavior remains valid for more complex systems.  

Our results show that the considered topological measures, both the simple ones (Betti numbers) and 
more complex ones (total persistence) appear very similar for the two and three dimensional suspension flows studied. 
Since the number of contacts and the geometry of particle packing are strongly dependent on the 
number of physical dimensions, one wonders whether there is some additional not yet discovered property of 
the interaction networks that leads to their dimension - independent properties.  

The interaction networks clearly play an important role in quantifying rheological response of sheared suspensions. 
While these networks are complex, particularly when considered for all interaction strengths at once, persistent homology
provides a way towards their analysis and quantification, removing a significant amount of complexity and isolating
rather straightforward measures that could be used to quantify the rheology.  Our results suggest that a similar type of
analysis may be of utility for the purpose of describing many other systems such that the interaction 
between the basic building blocks can be described in terms of interaction networks.
We also note a  potential in using persistent homology to analyze temporal evolution of evolving networks.  Well established measures for computing correlations
between networks at different times exist, and we expect that significant new 
insight could be reached by applying such measures and comparing them with 
evolving rheological properties.    

{\it Acknowledgments:}
The work on particle simulations carried out in this work was supported, in part, under National Science Foundation Grants CNS-0958379, CNS-0855217, ACI-1126113 and the City University of New York High Performance Computing Center at the College of Staten Island.
JFM was supported by NSF 160528.  LK was supported by NSF DMS-1521717 and ARO W911NF1810184.
MG was supported by FAPESP grants 2016/08704-5 and 2016/21032-6 and by CNPq grant 310740/2016-9, Brazil.
KM and MG were supported by NSF DMS-1521771, DMS-1622401, DMS-1839294 and DARPA contracts HR0011-16-2-0033 and FA8750-17-C-0054.
\bibliographystyle{unsrt}
\bibliography{dst,granulates}

\end{document}